\pdfoutput=1
\documentclass[11pt]{article}
\usepackage[letterpaper,margin=1in]{geometry}
\usepackage[T1]{fontenc}
\usepackage[utf8]{inputenc}
\usepackage{amsmath,mathtools}
\usepackage{newtxtext,newtxmath}
\usepackage{graphicx}
\usepackage{booktabs,longtable,array,calc}
\usepackage{microtype}
\usepackage{xcolor}
\usepackage{hyperref}
\usepackage{xurl}
\usepackage{caption}
\usepackage{enumitem}
\usepackage{etoolbox}
\usepackage{url}
\usepackage{fancyhdr}
\hypersetup{hidelinks,pdftitle={Aggregation Distortion in Multilevel Mediation: Estimand Geometry, Design Dependence, and Diagnostics},pdfauthor={Subir Hait}}
\providecommand{\tightlist}{\setlength{\itemsep}{0pt}\setlength{\parskip}{0pt}}
\providecommand{\real}[1]{#1}
\makeatletter
\patchcmd\longtable{\par}{\if@noskipsec\mbox{}\fi\par}{}{}
\makeatother

\begin{document}

% -----------------------------------------------------------------------------
% TITLE PAGE
% -----------------------------------------------------------------------------
\thispagestyle{fancy}
\begin{center}
\vspace*{1.35in}
{\LARGE\bfseries Aggregation Distortion in Multilevel Mediation:\par}
\vspace{0.18in}
{\Large\bfseries Estimand Geometry, Design Dependence, and Diagnostics\par}

\vspace{1.05in}
{\large Subir Hait\par}
\vspace{0.22in}
{\normalsize Department of Counseling, Educational Psychology, and Special Education\par}
{\normalsize Michigan State University\par}
\vspace{0.25in}
{\normalsize Correspondence: \texttt{haitsubi@msu.edu}\par}
{\normalsize ORCID: 0009-0004-9871-9677\par}
\end{center}
\clearpage
\setcounter{page}{2}

% -----------------------------------------------------------------------------
% ABSTRACT PAGE
% -----------------------------------------------------------------------------
\begin{abstract}
Researchers often summarize mediation in clustered data with a
single-level product-of-coefficients estimator. We show what this pooled
analysis estimates and why the target depends on the design. The pooled
\(a\)-path is weighted by the within--between variance composition of the
treatment. The pooled \(b\)-path uses a different weight, based on the
mediator variance that remains after residualizing on treatment and
covariates. Because these weights generally differ, the pooled indirect
effect can differ from a level-respecting combination of the within- and
between-level indirect effects. We derive the pooled probability limit
and split this difference into a nonlinearity term and a weight-mismatch
term. Equal \(b\)-paths remove the distortion, but equal \(a\)-paths do
not. Allowing the two weights to vary gives a bilinear surface with
saddle geometry, sharp corner bounds, and a zero-distortion frontier. We
also define a design-standardized indirect effect and practical
diagnostics. Simulations verify the identities and show that a common
decompose-first, cluster-weighted construction can overstate apparent
distortion. In an ECLS-K:2011 illustration with 8,363 children in 753
baseline schools, the within-level weights were 0.723 and 0.896, a
difference of -0.173 (95\% bootstrap CI {[}-0.193, -0.151{]}). The
conventional construction produced 1.76 times the absolute distortion
of the identity-consistent construction. The identity-consistent AD
interval included zero. Thus, the application shows unequal weighting
and sensitivity to construction, not evidence of nonzero population
distortion.
\end{abstract}


\clearpage

\section{1. Introduction}\label{introduction}

Mediation analysis separates a total effect into a direct part and an
indirect part that operates through an intermediate variable (Baron \&
Kenny, 1986; Imai, Keele, \& Tingley, 2010; MacKinnon, 2008;
VanderWeele, 2015). Many applications are clustered: students are nested
in schools, employees in firms, and patients in clinics. Yet mediation
is still often summarized with a single-level product of coefficients,
with clustering handled by a random intercept or a cluster-robust
standard error.

That approach treats clustering mainly as an inference problem. For
mediation, it can also change the estimand. The reason is simple: the
product of two weighted averages is generally not the weighted average
of two products. If the treatment--mediator and mediator--outcome paths
differ between and within clusters, the pooled product need not equal
the within-cluster indirect effect, the between-cluster indirect effect,
or a fixed mixture of the two.

Multilevel mediation work has long shown that within- and
between-cluster paths should be separated (Krull \& MacKinnon, 2001;
Preacher, Zyphur, \& Zhang, 2010; Zhang, Zyphur, \& Preacher, 2009).
Other work has clarified mediation under interference (VanderWeele,
Hong, Jones, \& Brown, 2013), interventional effects (Vansteelandt \&
Daniel, 2017), and the need to define the causal question before choosing
an estimator (Nguyen, Schmid, \& Stuart, 2021). The issue studied here is
different: what does the ordinary pooled product converge to, and how
does that quantity change when the design changes?

The answer depends on two weights. The pooled \(a\)-path combines the
within- and between-level \(a\)-paths according to the variance
composition of treatment. The pooled \(b\)-path uses the variance
composition of the mediator after residualizing on treatment and
covariates, by the Frisch--Waugh--Lovell theorem. There is no general
reason for those two weights to be equal.

From this difference, several results follow. The gap between the pooled
indirect effect and a level-respecting target separates exactly into a
nonlinearity component and a weight-mismatch component. The two
components behave differently: equal \(b\)-paths remove distortion,
whereas equal \(a\)-paths do not when the weights differ. If the two
weights are allowed to vary independently, the distortion forms a
bilinear surface over the unit square. This gives a zero-distortion
frontier, sharp corner bounds, and a simple way to study design
sensitivity. The same framework also leads to a design-standardized
indirect effect and connects directly to the random-coefficient
covariance correction of Kenny, Korchmaros, and Bolger (2003) and Bauer,
Preacher, and Gil (2006).

The paper is about estimands, not a new solution to causal
identification. The level-specific ignorability assumptions in Section
4 remain strong and untestable in observational data. When those
assumptions are not credible, the results should be read as statements
about linear projection parameters. The contribution is to show exactly
what pooling does to those parameters and how to diagnose the effect of
the design.

Sections 2--4 set up the problem and the level-specific estimands.
Sections 5--10 derive the pooled limit, its decomposition, and the
design geometry. Sections 11--14 cover random coefficients, estimation,
diagnostics, and simulation. Section 15 gives the ECLS-K illustration,
followed by implementation guidance and discussion in Sections 16--17.

\section{2. Positioning}\label{positioning}

The argument connects to three established literatures. The overlap is
useful, but the quantities studied here are not the same as the ones in
those literatures.

\subsection{2.1 Random-Coefficient Covariance
Corrections}\label{random-coefficient-covariance-corrections}

Kenny, Korchmaros, and Bolger (2003) and Bauer, Preacher, and Gil (2006)
showed that when the \(a\)- and \(b\)-paths vary across clusters, the
mean cluster-specific indirect effect contains a covariance term:
\[
\mathbb{E}\left\lbrack a_{j}b_{j} \right\rbrack
 = \bar{a}\bar{b} + \tau_{ab}.
\]
Corollary 1 below has a similar algebraic form, but it refers to a
different source of covariance.

The Kenny--Bauer term comes from genuine cluster-to-cluster variation in
a correctly specified multilevel model. Its expectation is taken over
the population of clusters. Our term comes from mixing fixed
within- and between-level paths in a pooled model. The relevant
two-point distribution is determined by variance weights from the
design. Thus, the Kenny--Bauer covariance is a population quantity to
estimate, whereas the covariance in Corollary 1 is a feature of what
pooling leaves out.

Both terms can appear in the same model. Proposition 8 gives the pooled
limit when random within-cluster paths and cross-level aggregation are
present together. The two covariance terms then enter separately.

\subsection{2.2 Sampling Error in Observed Cluster
Means}\label{sampling-error-in-observed-cluster-means}

Observed cluster means are noisy measures of latent cluster-level
constructs. Lüdtke et al.~(2008) and Preacher, Zhang, and Zyphur (2011)
showed how that noise can bias contextual effects and how latent
multilevel models can address it. Section 3 gives the corresponding
finite-cluster relation between the observed between-cluster variance
weight and the ICC.

Here the same sampling noise plays a different role. Our target is the
pooled estimator itself, and that estimator uses the variance structure
of the observed data. For this purpose, the observed weight is part of
the estimand rather than a nuisance to be removed. Section 4.3 treats
the latent target separately so that the two questions are not mixed.

\subsection{2.3 Ecological Inference and the Mundlak
Decomposition}\label{ecological-inference-and-the-mundlak-decomposition}

The ecological fallacy shows that group-level and individual-level
associations can differ (Robinson, 1950). It is also standard that a
pooled regression coefficient can be written as a variance-weighted
average of within- and between-group coefficients (Mundlak, 1978;
Raudenbush \& Bryk, 2002; Angrist \& Pischke, 2009). Lemma 2 uses that
result.

The new issue appears when two such coefficients are multiplied. A
single pooled coefficient is a convex combination and therefore lies
between its within and between counterparts. A pooled indirect effect
is bilinear. It can fall outside the range defined by \(IE_{W}\) and
\(IE_{B}\). That difference drives the results that follow.

\subsection{2.4 Main Contributions}\label{summary-of-contributions}

The paper develops the following results.

\begin{enumerate}
\def\labelenumi{\arabic{enumi}.}
\tightlist
\item
  It defines a within-level natural indirect effect and an observed-mean
  contextual between-level analogue under partial interference, with
  identification conditions and a separate treatment of the latent
  alternative (Section 4).
\item
  It gives an exact two-component decomposition of aggregation
  distortion and shows why homogeneity of the \(a\)- and \(b\)-paths
  have different implications (Theorem 2, Corollary 2).
\item
  It states the scope condition needed for the Mundlak identity under
  informative cluster size: residualize before decomposing and use
  person-weighted between-level projections (Proposition 4).
\item
  It derives the full two-weight distortion surface, including its
  bilinear geometry, saddle structure, and zero-distortion frontier
  (Theorem 3, Corollary 4).
\item
  It gives sharp global bounds on distortion from the four corners of
  the design square (Theorem 4).
\item
  It gives sharp bounds on the pooled estimand when the two design
  weights are known only to lie in intervals (Corollary 5).
\item
  It defines a design-standardized indirect effect and two diagnostics
  for design sensitivity (Theorem 6).
\item
  It combines cross-level aggregation with random-coefficient
  heterogeneity in one expression (Proposition 8).
\end{enumerate}

\section{3. Setup}\label{setup}

\subsection{3.1 Data Structure and the Latent--Observed
Distinction}\label{data-structure-and-the-latentobserved-distinction}

Let \(i = 1,\ldots,n_{j}\) index individuals and \(j = 1,\ldots,J\)
index clusters, with \(N = \sum_{j}^{}n_{j}\). For individual \(i\) in
cluster \(j\), let \(D_{ij}\) denote treatment, \(M_{ij}\) a mediator,
\(Y_{ij}\) an outcome, and \(\mathbf{X}_{ij}\) a covariate vector. Write
the treatment in latent components as

\[D_{ij} = \mu + B_{j} + W_{ij},\]

where \(B_{j}\) is the latent between-cluster component with
\(Var\left( B_{j} \right) = \sigma_{B}^{2}\), and \(W_{ij}\) is the
within-cluster deviation with
\(\mathbb{E}\left\lbrack W_{ij} \mid j \right\rbrack = 0\) and
\(Var\left( W_{ij} \right) = \sigma_{W}^{2}\).

Two between-cluster quantities must be distinguished. The \emph{latent}
between-cluster component is \(D_{j}^{B} = \mu + B_{j}\), with variance
\(\sigma_{B}^{2}\). The \emph{observed} cluster mean is
\({\bar{D}}_{j} = n_{j}^{- 1}\sum_{i}^{}D_{ij} = \mu + B_{j} + {\bar{W}}_{j}\),
with variance \(\sigma_{B}^{2} + \sigma_{W}^{2}/n_{j}\). The observed
within-cluster deviation is \(D_{ij}^{W} = D_{ij} - {\bar{D}}_{j}\).

The \emph{variance weights} are defined with respect to the observed
decomposition:

\[\omega_{B}(D) = \frac{Var\left( {\bar{D}}_{j} \right)}{Var\left( D_{ij} \right)},\quad\quad\omega_{W}(D) = \frac{Var\left( D_{ij}^{W} \right)}{Var\left( D_{ij} \right)}.\]

\textbf{Lemma 1 (Orthogonality of the Observed Decomposition; Relation
to the ICC).}

\emph{(a) (Orthogonality; exact under unequal cluster sizes.) For the
observed decomposition} \(D_{ij} = {\bar{D}}_{j} + D_{ij}^{W}\)\emph{,
the within-cluster deviations sum to zero inside every cluster by
construction, so for any cluster sizes} \(n_{j}\)\emph{,}

\[\sum_{j}^{}{\sum_{i}^{}{\bar{D}}_{j}}D_{ij}^{W} = \sum_{j}^{}{\bar{D}}_{j}\sum_{i}^{}\left( D_{ij} - {\bar{D}}_{j} \right) = 0,\]

\emph{and correspondingly}
\(Cov\left( {\bar{D}}_{j},D_{ij}^{W} \right) = 0\) \emph{in population.
Hence}
\(Var\left( D_{ij} \right) = Var\left( {\bar{D}}_{j} \right) + Var\left( D_{ij}^{W} \right)\)
\emph{and} \(\omega_{B}(D) + \omega_{W}(D) = 1\) \emph{exactly, whether
or not clusters are balanced.}

\emph{(b) (Relation to the ICC; exact under equal cluster sizes.) If}
\(n_{j} = n\) \emph{for all} \(j\)\emph{, then}
\(Var\left( {\bar{D}}_{j} \right) = \sigma_{B}^{2} + \sigma_{W}^{2}/n\)
\emph{and}
\(Var\left( D_{ij}^{W} \right) = \sigma_{W}^{2}(1 - 1/n)\)\emph{, so}

\[\omega_{B}(D) = ICC(D) + \frac{1 - ICC(D)}{n} > ICC(D),\]

\emph{where}
\(ICC(D) = \sigma_{B}^{2}/\left( \sigma_{B}^{2} + \sigma_{W}^{2} \right)\)\emph{.
The discrepancy vanishes as} \(n \rightarrow \infty\)\emph{.}

\emph{(c) (Unequal cluster sizes.) Under imbalance the relation in (b)
holds only approximately, with} \(n\) \emph{replaced by the harmonic
mean cluster size; the approximation degrades as the coefficient of
variation of cluster sizes grows.}

\emph{\textbf{Proof.}} Appendix B. \(\blacksquare\)

\textbf{Proposition 1 (Design-Effect Form of the Between Weight).}
\emph{For equal cluster sizes} \(n_{j} = n\)\emph{,}

\[\omega_{B}(D) = \frac{1 + (n - 1)ICC(D)}{n},\]

\emph{that is,} \(\omega_{B}(D)\) \emph{equals the design effect divided
by the cluster size. The discrepancy}
\(\delta = \omega_{B}(D) - ICC(D) = \left( 1 - ICC(D) \right)/n\)
\emph{is decreasing in both} \(n\) \emph{and} \(ICC(D)\)\emph{.}

\emph{\textbf{Proof.}} Appendix B. \(\blacksquare\)

\textbf{Remark 1.} Lemma 1 separates two issues that are easy to mix
up. The observed within--between decomposition is orthogonal for any
cluster sizes, so its variance weights always add to one. Balance is
needed only for the simple closed form that links \(\omega_{B}(D)\) to
the ICC. With unequal cluster sizes, the weights should be calculated
directly from the sample variances.

\textbf{Remark 2.} The between-cluster weight should not be replaced by
the ICC. For example, when \(ICC = 0.05\) and \(n = 10\),
\(\omega_{B}(D) = 0.145\), almost three times the ICC. Because this
weight enters the pooled estimand in Theorem 1, the substitution changes
the target itself, not just a descriptive variance summary.

\subsection{3.2 The Structural Model}\label{the-structural-model}

\textbf{Definition 1 (Two-Level Linear Mediation Model).} \emph{The
structural equations are}

\[M_{ij} = \alpha_{W}D_{ij}^{W} + \alpha_{B}{\bar{D}}_{j} + \mathbf{\gamma}_{M}^{\prime}\mathbf{X}_{ij} + \zeta_{j}^{M} + u_{ij},\]

\[Y_{ij} = \beta_{2W}M_{ij}^{W} + \beta_{2B}{\bar{M}}_{j} + \beta_{W}D_{ij}^{W} + \beta_{B}{\bar{D}}_{j} + \mathbf{\gamma}_{Y}^{\prime}\mathbf{X}_{ij} + \zeta_{j}^{Y} + e_{ij},\]

\emph{where} \(\zeta_{j}^{M} \sim \left( 0,\tau_{M}^{2} \right)\)
\emph{and} \(\zeta_{j}^{Y} \sim \left( 0,\tau_{Y}^{2} \right)\)
\emph{are cluster random effects and}
\(u_{ij} \sim \left( 0,\sigma_{M}^{2} \right)\)\emph{,}
\(e_{ij} \sim \left( 0,\sigma_{Y}^{2} \right)\) \emph{are
individual-level residuals, all mutually independent. The path
coefficients are constant across clusters; this is relaxed in Section
11.}

\textbf{Remark 3 (Observed projections).} Throughout the paper,
\(\alpha_{B}\) and \(\beta_{2B}\) refer to linear projections on the
\emph{observed} cluster means \({\bar{D}}_{j}\) and
\({\bar{M}}_{j}\). These are the between-level coefficients recovered by
the estimator in Section 13. If the target is instead a structural
coefficient on a latent cluster component, the estimand changes and is
generally not identified from observed means when cluster sizes remain
bounded. Section 4.3 treats that case. The two definitions agree only as
\(n_{j} \rightarrow \infty\).

We use the observed-projection convention because it matches the
quantity mixed by the pooled estimator and can be estimated without an
additional measurement-error model. It also keeps two problems
separate: aggregation across levels and unreliability of cluster means.

The model contains no treatment--mediator interaction and no random path
variation. Both restrictions are substantive and are revisited in
Sections 11 and 17.

\begin{figure}[htbp]
\centering
\includegraphics[width=0.80\textwidth]{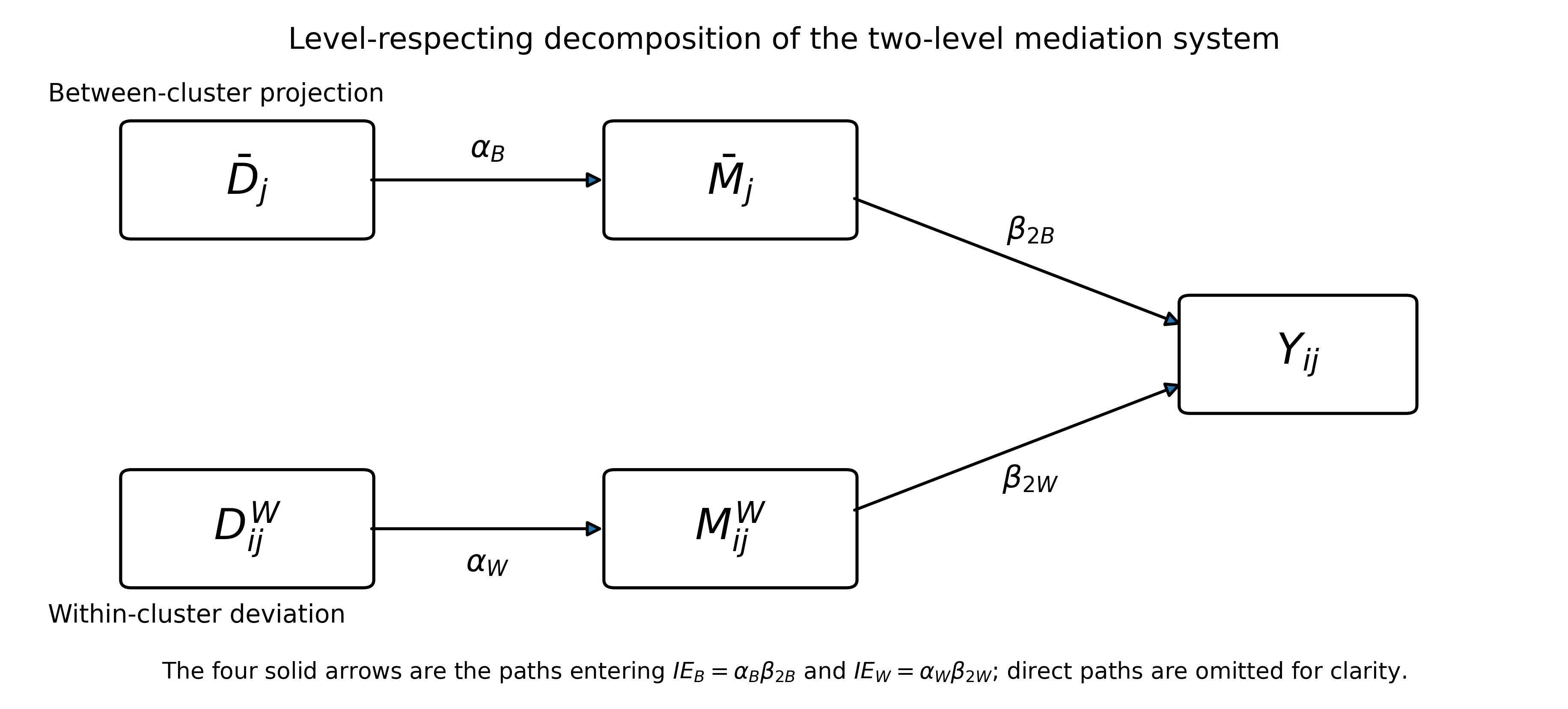}
\caption{Level-respecting representation of the two-level mediation system. The upper lane shows the observed between-cluster projection and the lower lane the within-cluster deviation. The four displayed paths define \(IE_B=\alpha_B\beta_{2B}\) and \(IE_W=\alpha_W\beta_{2W}\). Direct treatment-to-outcome paths are part of Definition 1 but are omitted here for clarity.}
\label{fig:decomposed-system}
\end{figure}

Figure~\ref{fig:decomposed-system} shows the decomposition used in the
paper. It starts with \(\bar D_j\) and \(D^W_{ij}\) as separate exposure
coordinates. We do not draw \(D_{ij}\rightarrow\bar D_j\), because the
cluster mean is a constructed summary rather than a downstream random
variable. The direct \(D\rightarrow Y\) paths are omitted so the four
paths that form the two indirect effects remain easy to see.

\section{4. Level-Specific Causal
Estimands}\label{level-specific-causal-estimands}

\subsection{4.1 Potential Outcomes Under Partial
Interference}\label{potential-outcomes-under-partial-interference}

Because treatment has both within- and between-cluster components, a
unit's outcome may depend on the treatment pattern in its own cluster.
We use partial interference: assignments in other clusters do not
matter, and the own-cluster treatment pattern enters through its mean.
Potential outcomes are indexed by \((b,w)\), where \(b\) is the realized
cluster mean and \(w\) is the individual's deviation from that mean, so
\(D_{ij}=b+w\).

Let \(M_{ij}(b,w)\) denote the mediator under cluster-mean exposure
\(b\) and individual deviation \(w\), and \(Y_{ij}(b,w,m)\) the outcome
under that exposure pair with the mediator set to \(m\). Consistency
requires \(M_{ij} = M_{ij}\left( {\bar{D}}_{j},D_{ij}^{W} \right)\) and
\(Y_{ij} = Y_{ij}\left( {\bar{D}}_{j},D_{ij}^{W},M_{ij} \right)\).

\textbf{Definition 2 (Within-Level Natural Effects and Between-Level
Contextual Analogues).} \emph{For a unit shift in each treatment
component,}

\[NIE_{W} = \mathbb{E}\left\lbrack Y\left( b,w + 1,M(b,w + 1) \right) - Y\left( b,w + 1,M(b,w) \right) \right\rbrack,\]

\[NDE_{W} = \mathbb{E}\left\lbrack Y\left( b,w + 1,M(b,w) \right) - Y\left( b,w,M(b,w) \right) \right\rbrack,\]

\[NIE_{B} = \mathbb{E}\left\lbrack Y\left( b + 1,w,M(b + 1,w) \right) - Y\left( b + 1,w,M(b,w) \right) \right\rbrack,\]

\[NDE_{B} = \mathbb{E}\left\lbrack Y\left( b + 1,w,M(b,w) \right) - Y\left( b,w,M(b,w) \right) \right\rbrack.\]

The within-level contrast changes an individual's position relative to
the cluster while holding the realized cluster mean fixed. This is the
usual type of natural indirect effect. The between-level contrast changes
the realized cluster mean while holding the individual's relative
position fixed. We call it a \emph{contextual analogue} because the
observed cluster mean includes the focal individual and is jointly
determined by the cluster members. In applications where that shift has
no clear intervention interpretation, \(NIE_{B}\) should be read as a
descriptive contextual decomposition.

\subsection{4.2 Identification}\label{identification}

\textbf{Assumption 1 (Level-Specific Sequential Ignorability).}
\emph{Conditional on} \(\mathbf{X}_{ij}\) \emph{and any cluster-level
covariates: (W1)} \(\{ Y(b,w^{\prime},m),M(b,w)\}\perp\!\!\!\perp W_{ij}\)\emph{;
(W2)} \(Y(b,w,m)\perp\!\!\!\perp M_{ij} \mid W_{ij},{\bar{D}}_{j}\)\emph{; (B1)}
\(\{ Y(b^{\prime},w,m),M(b,w)\}\perp\!\!\!\perp{\bar{D}}_{j}\)\emph{; (B2)}
\(Y(b,w,m)\perp\!\!\!\perp{\bar{M}}_{j} \mid {\bar{D}}_{j}\)\emph{. The pairs
(W1)--(W2) and (B1)--(B2) are logically independent.}

\textbf{Assumption 2 (Partial Interference and Positivity).}
\emph{Potential outcomes for units in cluster} \(j\) \emph{do not depend
on assignments in clusters} \(j^{\prime} \neq j\)\emph{, and depend on
within-cluster assignments only through}
\(\left( {\bar{D}}_{j},D_{ij}^{W} \right)\)\emph{. Each level-specific
exposure component has positive variance.}

\textbf{Proposition 2 (Identification Under the Linear Model).}
\emph{Under Definition 1, Assumptions 1 and 2, and no
treatment--mediator interaction at either level,}

\[NIE_{W} = \alpha_{W}\beta_{2W},\quad\quad NIE_{B} = \alpha_{B}\beta_{2B},\quad\quad NDE_{W} = \beta_{W},\quad\quad NDE_{B} = \beta_{B}.\]

\emph{\textbf{Proof.}} Appendix B. \(\blacksquare\)

\subsection{4.3 Latent Between-Level Mechanisms and Measurement
Error}\label{latent-between-level-mechanisms-and-measurement-error}

An analyst may instead care about a \emph{latent} between-cluster
mechanism that removes sampling noise from the observed cluster means.
Write those paths as \(\alpha_{B}^{\dagger}\) and
\(\beta_{2B}^{\dagger}\). That is a valid target, but it is different
from the observed-mean projection estimated by a cluster-means
regression.

\textbf{Proposition 3 (Attenuation of the Between-Level}
\(a\)\textbf{-Path).} \emph{Suppose the mediator equation is written on
latent components,}
\(M_{ij} = \alpha_{W}W_{ij} + \alpha_{B}^{\dagger}B_{j} + \zeta_{j}^{M} + u_{ij}\)\emph{,
with} \(B_{j}\bot W_{ij}\) \emph{and equal cluster sizes} \(n\)\emph{.
Then the between-cluster regression of} \({\bar{M}}_{j}\) \emph{on}
\({\bar{D}}_{j}\) \emph{satisfies}

\[{\widehat{\alpha}}_{B}\overset{p}{\rightarrow}\lambda_{n}\,\alpha_{B}^{\dagger},\quad\quad\lambda_{n} = \frac{\sigma_{B}^{2}}{\sigma_{B}^{2} + \sigma_{W}^{2}/n} = \frac{n\, ICC(D)}{1 + (n - 1)ICC(D)},\]

\emph{where} \(\lambda_{n}\) \emph{is the reliability of the observed
cluster mean. The attenuation persists under} \(J \rightarrow \infty\)
\emph{with} \(n\) \emph{fixed and vanishes only as}
\(n \rightarrow \infty\)\emph{.}

\emph{\textbf{Proof.}} Appendix B. \(\blacksquare\)

\textbf{Remark 4 (The} \(b\)\textbf{-path is not a scalar reliability
problem).} Proposition 3 applies cleanly to the \(a\)-path because the
between-level regression has one error-prone regressor. The \(b\)-path
has two: \({\bar{M}}_{j}\) and \({\bar{D}}_{j}\). Their measurement
errors, \({\bar{u}}_{j}\) and \({\bar{W}}_{j}\), can also be related
through \(\alpha_{W}\). With two mismeasured regressors, the bias acts on
the coefficient vector as a matrix problem. There is no general scalar
reliability correction, and the coefficient need not be attenuated
toward zero.

For that reason, we do not treat the latent indirect effect as being
attenuated by a product of two reliabilities. Such a result needs extra
assumptions, for example independent classical measurement errors or a
residualization that reduces the problem to one orthogonal error-prone
regressor at a time. Under those conditions an approximate
\(\lambda_{n}^{D}\lambda_{n}^{M}\) correction may be justified.
Otherwise the multivariate errors-in-variables analysis of Lüdtke et
al.~(2008) is the relevant framework, and the direction of bias must be
derived for the particular model.

Theorem 7 therefore concerns the observed-projection estimands in
Remark 3, not latent structural paths. If the latent mechanism is the
target, a multilevel latent covariate model (Lüdtke et al., 2008) or
latent-aggregation SEM (Preacher, Zhang, \& Zyphur, 2011) is more
appropriate than a post hoc reliability adjustment; Lüdtke, Marsh,
Robitzsch, and Trautwein (2011) compare these approaches. The
aggregation results in Sections 5--11 still apply algebraically once a
set of level-specific paths has been defined. Cluster-mean
unreliability and cross-level aggregation are separate problems and can
occur together.

For the rest of the paper, write \(IE_{W}=\alpha_{W}\beta_{2W}\) and
\(IE_{B}=\alpha_{B}\beta_{2B}\). The algebra below does not require
Assumption 1. If causal identification is not credible, the results
still describe how the pooled and decomposed projection estimands are
related.

\section{5. The Pooled Estimand}\label{the-pooled-estimand}

\subsection{5.1 Two Weight Systems}\label{two-weight-systems}

An analyst who ignores the hierarchy fits

\[M_{ij} = \alpha_{0} + \alpha_{\text{pool}}D_{ij} + \varepsilon_{ij}^{M},\quad\quad Y_{ij} = \beta_{0} + \beta_{\text{pool}}D_{ij} + \beta_{2,\text{pool}}M_{ij} + \varepsilon_{ij}^{Y},\]

and reports
\(IE_{\text{pool}} = {\widehat{\alpha}}_{\text{pool}}{\widehat{\beta}}_{2,\text{pool}}\).

\textbf{Lemma 2 (Mundlak Representation of Pooled Coefficients).}
\emph{Under Definition 1 and condition (A6) of Appendix A,}

\[\alpha_{\text{pool}}\overset{p}{\rightarrow}\omega_{W}(D)\alpha_{W} + \omega_{B}(D)\alpha_{B},\quad\quad\beta_{2,\text{pool}}\overset{p}{\rightarrow}\omega_{W}(M \mid D)\beta_{2W} + \omega_{B}(M \mid D)\beta_{2B},\]

\emph{where} \(\omega_{W}(M \mid D)\) \emph{and}
\(\omega_{B}(M \mid D)\) \emph{are the variance weights of the mediator
after residualization on} \(D\) \emph{and the covariates, and}
\(\alpha_{B}\)\emph{,} \(\beta_{2B}\) \emph{are the person-weighted
between-level projections of Proposition 4. In general}
\(\omega_{B}(M \mid D) \neq \omega_{B}(D)\)\emph{.}

\emph{\textbf{Proof.}} Appendix B. \(\blacksquare\)

\textbf{Remark 5 (Ordering with covariates).} With covariates, the
order of operations matters. First residualize \(D\), \(M\), and \(Y\)
on \(\mathbf{X}\) at the individual level. Then form the within- and
between-cluster components and compute the weights and paths from those
residuals. If the variables are decomposed first and covariates are
adjusted separately within levels, the resulting coefficients need not
satisfy Lemma 2. Section 15.1 shows the size of this difference in the
ECLS-K example.

The key point is that the two weights generally differ. By the
Frisch--Waugh--Lovell theorem, the pooled \(b\)-path depends on the
mediator after residualization. That residualization can remove different
amounts of variance at the within and between levels. Section 14.3 shows
that nuisance variance in the mediator model alone can therefore move
the \(b\)-path weight while leaving the structural paths unchanged.

\textbf{Proposition 4 (Person- Versus Cluster-Weighted Between-Level
Paths).} \emph{Lemma 2 holds when} \(\alpha_{B}\) \emph{and}
\(\beta_{2B}\) \emph{are the \textbf{person-weighted} between-level
projections, that is, the cluster-means regressions weighted by}
\(n_{j}\)\emph{. Write} \(\alpha_{B}^{per}\) \emph{for that coefficient
and} \(\alpha_{B}^{clu}\) \emph{for the unweighted cluster-means
coefficient in which each cluster contributes one observation. Then:}

\emph{(a) If} \(n_{j} = n\) \emph{for all} \(j\)\emph{,}
\(\alpha_{B}^{per} = \alpha_{B}^{clu}\) \emph{and the distinction is
vacuous.}

\emph{(b) In general} \(\alpha_{B}^{per} \neq \alpha_{B}^{clu}\)\emph{,
and}

\[\alpha_{B}^{per} - \alpha_{B}^{clu} = \frac{{Cov}_{j}\left( n_{j},\mspace{6mu}\left( {\bar{D}}_{j} - \bar{D} \right)\left( {\bar{M}}_{j} - \alpha_{B}^{clu}{\bar{D}}_{j} \right) \right)}{\sum_{j}^{}n_{j}\left( {\bar{D}}_{j} - {\bar{D}}_{per} \right)^{2}/\sum_{j}^{}n_{j}},\]

\emph{so the two coincide whenever cluster size is uncorrelated with the
cluster-level regression residual and with the cluster-level exposure.
Otherwise the pooled estimator mixes} \(\alpha_{B}^{per}\)\emph{, not}
\(\alpha_{B}^{clu}\)\emph{, and substituting the latter into Theorem 1
leaves a residual of the same order as} \(A\) \emph{itself.}

\emph{\textbf{Proof.}} Appendix B. \(\blacksquare\)

The person- and cluster-weighted projections answer different questions.
The first is the between-level quantity that the pooled individual-level
regression mixes. The second gives each cluster equal weight and may be a
reasonable contextual estimand in its own right. Under informative
cluster size, however, the two need not agree. This distinction is well
known in cluster trials (Seaman, Pavlou, \& Copas, 2014; Kahan, Li,
Copas, \& Harhay, 2023). Section 15 shows why it also matters for
mediation: the two between-level indirect-effect products are close, but
the cluster-weighted paths miss the fitted pooled \(a\)- and \(b\)-paths
by 3.79\% and -2.08\%.

\textbf{Definition 3 (Weight Discrepancy).}
\(\Delta = \omega_{B}(M \mid D) - \omega_{B}(D)\).

\subsection{5.2 The Pooled Limit and the Decomposed
Target}\label{the-pooled-limit-and-the-decomposed-target}

\textbf{Theorem 1 (Pooled Probability Limit).} \emph{Under Definition 1
and Lemma 2,}

\[plim\left( IE_{\text{pool}} \right) = \left\lbrack \omega_{W}(D)\alpha_{W} + \omega_{B}(D)\alpha_{B} \right\rbrack\left\lbrack \omega_{W}(M \mid D)\beta_{2W} + \omega_{B}(M \mid D)\beta_{2B} \right\rbrack.\]

\emph{\textbf{Proof.}} Appendix B. \(\blacksquare\)

To measure the effect of pooling, we compare it with

\[IE_{\text{dec}} = \omega_{W}(D)\, IE_{W} + \omega_{B}(D)\, IE_{B}.\]

This target uses the same treatment weights as the pooled \(a\)-path and
can be defined before the outcome model is fitted. We use
\(IE_{\text{dec}}\) as a reference for measuring aggregation, not as the
preferred substantive estimand. It still depends on the realized
design, which motivates the standardized effect in Section 10.

\textbf{Definition 4 (Aggregation Distortion).}
\(A = plim\left( IE_{\text{pool}} \right) - IE_{\text{dec}}\).

We call \(A\) \emph{distortion} rather than \emph{bias} because it is a
difference between two estimands, not an estimation error relative to a
single fixed parameter.

\section{6. The Structure of Aggregation
Distortion}\label{the-structure-of-aggregation-distortion}

\subsection{6.1 The Two-Component
Decomposition}\label{the-two-component-decomposition}

\textbf{Theorem 2 (Exact Two-Component Decomposition).} \emph{Let}
\(\bar{\alpha} = \omega_{W}(D)\alpha_{W} + \omega_{B}(D)\alpha_{B}\)
\emph{and} \(\Delta = \omega_{B}(M \mid D) - \omega_{B}(D)\)\emph{.
Then}

\[A = \underset{A_{nl}\ \left( \text{nonlinearity} \right)}{\underbrace{- \,\omega_{W}(D)\,\omega_{B}(D)\,\left( \alpha_{W} - \alpha_{B} \right)\left( \beta_{2W} - \beta_{2B} \right)}}\mspace{6mu} + \mspace{6mu}\underset{A_{wt}\ \left( \text{weight mismatch} \right)}{\underbrace{- \,\bar{\alpha}\,\Delta\,\left( \beta_{2W} - \beta_{2B} \right)}},\]

\emph{equivalently}

\[A = - \,\left( \beta_{2W} - \beta_{2B} \right)\left\lbrack \omega_{W}(D)\omega_{B}(D)\left( \alpha_{W} - \alpha_{B} \right) + \bar{\alpha}\Delta \right\rbrack.\]

\emph{\textbf{Proof.}} Appendix B. \(\blacksquare\)

\(A_{nl}\) is the bilinearity effect: it would be present even if both
equations weighted the levels identically. \(A_{wt}\) exists only
because the two equations weight the levels differently.

\textbf{Corollary 1 (Covariance Representation).} \emph{Let} \(L\)
\emph{be a random level index with}
\(\mathbb{P}(L = W) = \omega_{W}(D)\) \emph{and}
\(\mathbb{P}(L = B) = \omega_{B}(D)\)\emph{. Then}
\(A_{nl} = - {Cov}_{\omega}\left( \alpha_{L},\beta_{L} \right)\)\emph{,
so that under} \(\Delta = 0\)\emph{,}
\(A = - {Cov}_{\omega}\left( \alpha_{L},\beta_{L} \right)\)
\emph{exactly: pooling discards precisely the design-weighted covariance
between where treatment moves the mediator and where the mediator moves
the outcome.}

\emph{\textbf{Proof.}} Appendix B. \(\blacksquare\)

\textbf{Corollary 2 (Asymmetry of the Homogeneity Conditions).}
\emph{Under} \(\Delta \neq 0\) \emph{and} \(\bar{\alpha} \neq 0\)\emph{:
(a)} \(\beta_{2W} = \beta_{2B}\) \emph{implies} \(A = 0\)\emph{, for any
weights and any degree of} \(a\)\emph{-path heterogeneity; (b)}
\(\alpha_{W} = \alpha_{B}\) \emph{does not imply} \(A = 0\) \emph{--- in
that case} \(A_{nl} = 0\) \emph{but}
\(A = A_{wt} = - \alpha\Delta\left( \beta_{2W} - \beta_{2B} \right)\)\emph{,
nonzero whenever} \(\beta_{2W} \neq \beta_{2B}\)\emph{.}

\emph{\textbf{Proof.}} Immediate from the factorization in Theorem 2.
\(\blacksquare\)

\textbf{Remark 6.} Under equal weights, the formula looks symmetric in
the \(a\)- and \(b\)-paths. Theorem 2 shows that this symmetry disappears
once the two equations use different weights. Heterogeneity in the
\(b\)-path is then necessary for distortion because that path is mixed
once by \(\omega(M\mid D)\) in the pooled model and again by
\(\omega(D)\) in the decomposed target. The \(a\)-path is not exposed to
that same mismatch.

\subsection{6.2 Sign, Magnitude, and Component
Dominance}\label{sign-magnitude-and-component-dominance}

\textbf{Proposition 5 (Sign and Magnitude).} \emph{Assume}
\(\omega_{W}(D),\omega_{B}(D) > 0\)\emph{.}

\emph{(a)}
\(sign\left( A_{nl} \right) = - sign\left( \alpha_{W} - \alpha_{B} \right)sign\left( \beta_{2W} - \beta_{2B} \right)\)\emph{.
When both path differences share a sign, pooling understates the
decomposed effect; when they differ, pooling overstates it.}

\emph{(b)}
\(\left| A_{nl} \right| \leq \frac{1}{4}\left| \alpha_{W} - \alpha_{B} \right|\,\left| \beta_{2W} - \beta_{2B} \right|\)\emph{,
with equality at} \(\omega_{W}(D) = 1/2\)\emph{.}

\emph{(c) Both components are bounded for fixed paths, since}
\(|\Delta| \leq 1\) \emph{gives}
\(\left| A_{wt} \right| \leq \left| \bar{\alpha} \right|\left| \beta_{2W} - \beta_{2B} \right|\)\emph{.
What is unbounded is their ratio:}

\[\frac{\left| A_{wt} \right|}{\left| A_{nl} \right|} = \frac{\left| \bar{\alpha} \right|\,|\Delta|}{\omega_{W}(D)\,\omega_{B}(D)\,\left| \alpha_{W} - \alpha_{B} \right|} \rightarrow \infty\quad\text{as }\left| \alpha_{W} - \alpha_{B} \right| \rightarrow 0.\]

\emph{Since} \(\omega_{W}(D)\omega_{B}(D) \leq 1/4\)\emph{, the ratio
exceeds one whenever}
\(\left| \bar{\alpha}\Delta \right| > \frac{1}{4}\left| \alpha_{W} - \alpha_{B} \right|\)\emph{.}

\emph{(d) Cancellation:} \(A = 0\) \emph{with both components nonzero
whenever}
\(\Delta = - \omega_{W}(D)\omega_{B}(D)\left( \alpha_{W} - \alpha_{B} \right)/\bar{\alpha}\)\emph{.}

\emph{\textbf{Proof.}} Appendix B. \(\blacksquare\)

Part (c) explains why the weight-mismatch term can matter even when it is
not large in absolute value. The nonlinearity term is limited by the
product of the two path gaps, whereas the weight-mismatch term depends on
the pooled \(a\)-path and the \(b\)-path gap. If the \(a\)-path differs
little across levels, \(A_{nl}\) can be very small and \(A_{wt}\) can
dominate the total.

\textbf{Remark 7.} A small observed \(A\) does not imply that pooling is
generally harmless. The two components can cancel. A replication with a
different variance composition can therefore move away from zero even if
the underlying paths are unchanged. Section 7 describes this as movement
away from the zero-distortion frontier.

\subsection{\texorpdfstring{6.3 Generalization to \(K\)
Levels}{6.3 Generalization to K Levels}}\label{generalization-to-k-levels}

\textbf{Corollary 3 (}\(K\)\textbf{-Level Generalization).} \emph{With
level-specific paths}
\(\left( \alpha_{\ell},\beta_{\ell} \right)\)\emph{, treatment weights}
\(\omega_{\ell}\) \emph{summing to one, and mediator weights}
\(\omega_{\ell}^{M}\)\emph{,}

\[A_{K} = - {Cov}_{\omega}\left( \alpha_{\ell},\beta_{\ell} \right)\mspace{6mu} - \mspace{6mu}\bar{\alpha}\sum_{\ell = 1}^{K}\left( \omega_{\ell}^{M} - \omega_{\ell} \right)\beta_{\ell},\]

\emph{where}
\(\bar{\alpha} = \sum_{\ell}^{}\omega_{\ell}\alpha_{\ell}\)\emph{.}

\emph{\textbf{Proof.}} Appendix B. \(\blacksquare\)

\subsection{6.4 A Worked Numerical
Example}\label{a-worked-numerical-example}

Take \(\alpha_{W} = 0.30\), \(\alpha_{B} = 0.50\),
\(\beta_{2W} = 0.40\), \(\beta_{2B} = 0.70\), with
\(\omega_{W}(D) = 0.70\) and \(\omega_{W}(M \mid D) = 0.65\), so
\(\Delta = + 0.05\). Then \(\bar{\alpha} = 0.360\) and the pooled
\(b\)-path limit is \(0.505\), giving
\(plim\left( IE_{\text{pool}} \right) = 0.18180\) against a decomposed
target of \(0.18900\), so \(A = - 0.00720\).

Theorem 2 recovers this from the paths alone. With
\(\Delta_{\alpha} = - 0.20\) and \(\Delta_{\beta} = - 0.30\),

\[A_{nl} = - (0.70)(0.30)( - 0.20)( - 0.30) = - 0.01260,\quad\quad A_{wt} = - (0.360)(0.05)( - 0.30) = + 0.00540,\]

The two components sum to \(-0.00720\). They have opposite signs, so the
weight-mismatch term offsets 43\% of the nonlinearity term. Using only
the equal-weight expression would give \(-0.01260\), overstating the
distortion by 75\%.

The asymmetry can be seen directly. If
\(\beta_{2B}=\beta_{2W}\), both components are zero regardless of the
\(a\)-path gap. If instead \(\alpha_{B}=\alpha_{W}=0.30\),
\(A_{nl}=0\) but \(A_{wt}=+0.00450\).

\section{7. The Two-Weight Design
Surface}\label{the-two-weight-design-surface}

Sections 5 and 6 evaluate distortion at the observed pair of weights.
Those weights can change separately. The treatment weight changes with
the cluster-size distribution and the allocation of treatment variance
across levels. The residualized-mediator weight also depends on nuisance
variance in the mediator model. It is therefore useful to examine the
full two-dimensional design space.

\subsection{7.1 The Design Square}\label{the-design-square}

\textbf{Definition 5 (Design Coordinates and the Distortion Surface).}
\emph{Let}

\[p = \omega_{W}(D) \in \lbrack 0,1\rbrack,\quad\quad q = \omega_{W}(M \mid D) \in \lbrack 0,1\rbrack,\]

\emph{and define the level-mixed paths and the pooled and decomposed
estimands as functions of the design:}

\[a(p) = \alpha_{B} + p\,\Delta_{\alpha},\quad\quad b(q) = \beta_{2B} + q\,\Delta_{\beta},\quad\quad\Delta_{\alpha} = \alpha_{W} - \alpha_{B},\quad\Delta_{\beta} = \beta_{2W} - \beta_{2B},\]

\[P(p,q) = a(p)\, b(q),\quad\quad T(p) = p\, IE_{W} + (1 - p)\, IE_{B},\quad\quad A(p,q) = P(p,q) - T(p).\]

\emph{The unit square} \(\lbrack 0,1\rbrack^{2}\) \emph{is the design
space; the diagonal} \(p = q\) \emph{is the equal-weight line on which
the textbook closed form is exact, and the realized design occupies the
point} \(\left( \widehat{p},\widehat{q} \right)\) \emph{with}
\(\widehat{q} - \widehat{p} = - \Delta\)\emph{.}

\subsection{7.2 Bilinear Geometry}\label{bilinear-geometry}

\textbf{Theorem 3 (Bilinear Design Geometry).} \(P\) \emph{and} \(A\)
\emph{are bilinear on} \(\lbrack 0,1\rbrack^{2}\)\emph{: affine in}
\(p\) \emph{for fixed} \(q\) \emph{and affine in} \(q\) \emph{for fixed}
\(p\)\emph{. Their derivatives are}

\[\frac{\partial P}{\partial p} = \Delta_{\alpha}\, b(q),\quad\quad\frac{\partial P}{\partial q} = \Delta_{\beta}\, a(p),\quad\quad\frac{\partial^{2}P}{\partial p\,\partial q} = \Delta_{\alpha}\Delta_{\beta},\quad\quad\frac{\partial^{2}P}{\partial p^{2}} = \frac{\partial^{2}P}{\partial q^{2}} = 0,\]

\emph{and, since} \(T\) \emph{does not depend on} \(q\) \emph{and is
linear in} \(p\)\emph{,}

\[\frac{\partial A}{\partial p} = \Delta_{\alpha}\, b(q) - \left( IE_{W} - IE_{B} \right),\quad\quad\frac{\partial A}{\partial q} = \Delta_{\beta}\, a(p),\quad\quad\frac{\partial^{2}A}{\partial p\,\partial q} = \Delta_{\alpha}\Delta_{\beta}.\]

\emph{Consequently the Hessian of either surface is the hollow matrix}

\[\mathbf{H} = \begin{pmatrix}
0 & \Delta_{\alpha}\Delta_{\beta} \\
\Delta_{\alpha}\Delta_{\beta} & 0
\end{pmatrix},\quad\quad\text{eigenvalues } \pm \left| \Delta_{\alpha}\Delta_{\beta} \right|,\quad tr\mathbf{H} = 0.\]

\emph{The entire curvature of both surfaces is therefore cross-weight
curvature: neither weight generates curvature on its own, the surfaces
are ruled --- straight lines along every axis-parallel section --- and
every non-degenerate stationary point is a saddle. Equal-and-opposite
eigenvalues summing to zero are the empirical signature of the result,
and are directly checkable in any application.}

\emph{\textbf{Proof.}} Appendix B. \(\blacksquare\)

The geometry makes the source of distortion clear. Holding either weight
fixed leaves a linear relationship in the other weight. Curvature appears
only when both weights move. Its magnitude is
\(\Delta_{\alpha}\Delta_{\beta}\), the product of the two level
differences in the paths.

\textbf{Corollary 4 (Zero-Distortion Frontier).} \emph{Suppose}
\(\beta_{2W} \neq \beta_{2B}\)\emph{. For any} \(p\) \emph{with}
\(a(p) \neq 0\)\emph{, the unique} \(q\) \emph{at which pooling and the
decomposed target agree is}

\[q_{0}(p) = \frac{T(p)/a(p) - \beta_{2B}}{\beta_{2W} - \beta_{2B}}.\]

\emph{The set}
\(\mathcal{Z} = \{\left( p,q_{0}(p) \right):q_{0}(p) \in \lbrack 0,1\rbrack\}\)
\emph{is the zero-distortion frontier. It contains the corners}
\((0,0)\) \emph{and} \((1,1)\)\emph{, and it partitions the design
square into a region where pooling overstates the decomposed effect and
a region where it understates. If} \(\beta_{2W} = \beta_{2B}\)
\emph{then} \(A \equiv 0\) \emph{on the whole square and}
\(\mathcal{Z} = \lbrack 0,1\rbrack^{2}\)\emph{.}

\emph{\textbf{Proof.}} Appendix B. \(\blacksquare\)

\textbf{Remark 8.} The frontier also shows why the two homogeneity
conditions differ. If the \(b\)-paths are equal,
\(\mathcal{Z}=[0,1]^2\), so distortion is zero for every design. If only
the \(a\)-paths are equal, \(a(p)\) is constant and the zero set becomes
a horizontal line. Designs away from that line still distort. Thus,
observing \(A\approx0\) says that the realized design lies near the
frontier; it does not show that the underlying mechanism is
design-insensitive.

\subsection{7.3 Sharp Global Bounds}\label{sharp-global-bounds}

\textbf{Theorem 4 (Sharp Global Distortion Bounds).} \emph{Because}
\(A\) \emph{is bilinear on a rectangle, its extrema over}
\(\lbrack 0,1\rbrack^{2}\) \emph{are attained at corners. The corner
values are}

\[A(0,0) = 0,\quad\quad A(1,1) = 0,\quad\quad A(0,1) = \alpha_{B}\,\Delta_{\beta},\quad\quad A(1,0) = - \alpha_{W}\,\Delta_{\beta},\]

\emph{and therefore}

\[\sup_{(p,q) \in \lbrack 0,1\rbrack^{2}}\left| A(p,q) \right|\mspace{6mu} = \mspace{6mu}\left| \beta_{2W} - \beta_{2B} \right| \cdot \max\left( \left| \alpha_{W} \right|,\,\left| \alpha_{B} \right| \right).\]

\emph{\textbf{Proof.}} Appendix B. \(\blacksquare\)

The bound has a simple interpretation. It is zero exactly when the
\(b\)-paths agree. It depends on the size of the \(a\)-path itself, not
only on the difference between the two \(a\)-paths. The largest
distortion occurs at \((0,1)\) or \((1,0)\), where the two regression
equations put their weight on opposite levels. Designs can approach this
case when little between-cluster mediator variance remains after
adjustment for treatment, pushing \(q\) toward one while \(p\) remains
moderate.

\subsection{7.4 Partial Identification Under Uncertain Design
Weights}\label{partial-identification-under-uncertain-design-weights}

For a replication or a target population, the exact design weights may be
unknown. Often, however, plausible ranges for \(p\) and \(q\) can be
specified.

\textbf{Corollary 5 (Design-Robustness Region).} \emph{Suppose only
that} \(p \in \left\lbrack p_{L},p_{U} \right\rbrack\) \emph{and}
\(q \in \left\lbrack q_{L},q_{U} \right\rbrack\)\emph{. Because} \(P\)
\emph{is bilinear, its extrema over that rectangle are attained at its
four corners, so}

\[\mathcal{R} = \left\lbrack \min_{(p,q) \in \mathcal{C}}P(p,q),\mspace{6mu}\max_{(p,q) \in \mathcal{C}}P(p,q) \right\rbrack,\quad\quad\mathcal{C} = \{ p_{L},p_{U}\} \times \{ q_{L},q_{U}\},\]

\emph{are sharp bounds on the pooled estimand, attained and not merely
valid. The same holds for} \(A\)\emph{. If} \(\mathcal{R}\)
\emph{excludes zero, the sign of the pooled indirect effect is invariant
over all admissible designs; if} \(0 \in \mathcal{R}\)\emph{, the sign
is design-fragile.}

\emph{\textbf{Proof.}} Appendix B. \(\blacksquare\)

This gives a simple robustness check: evaluate the four corners of the
specified design rectangle. The same idea extends the one-dimensional
suppression diagnostic in Section 9 to uncertainty in both weights.

\begin{figure}[htbp]
\centering
\includegraphics[width=\textwidth]{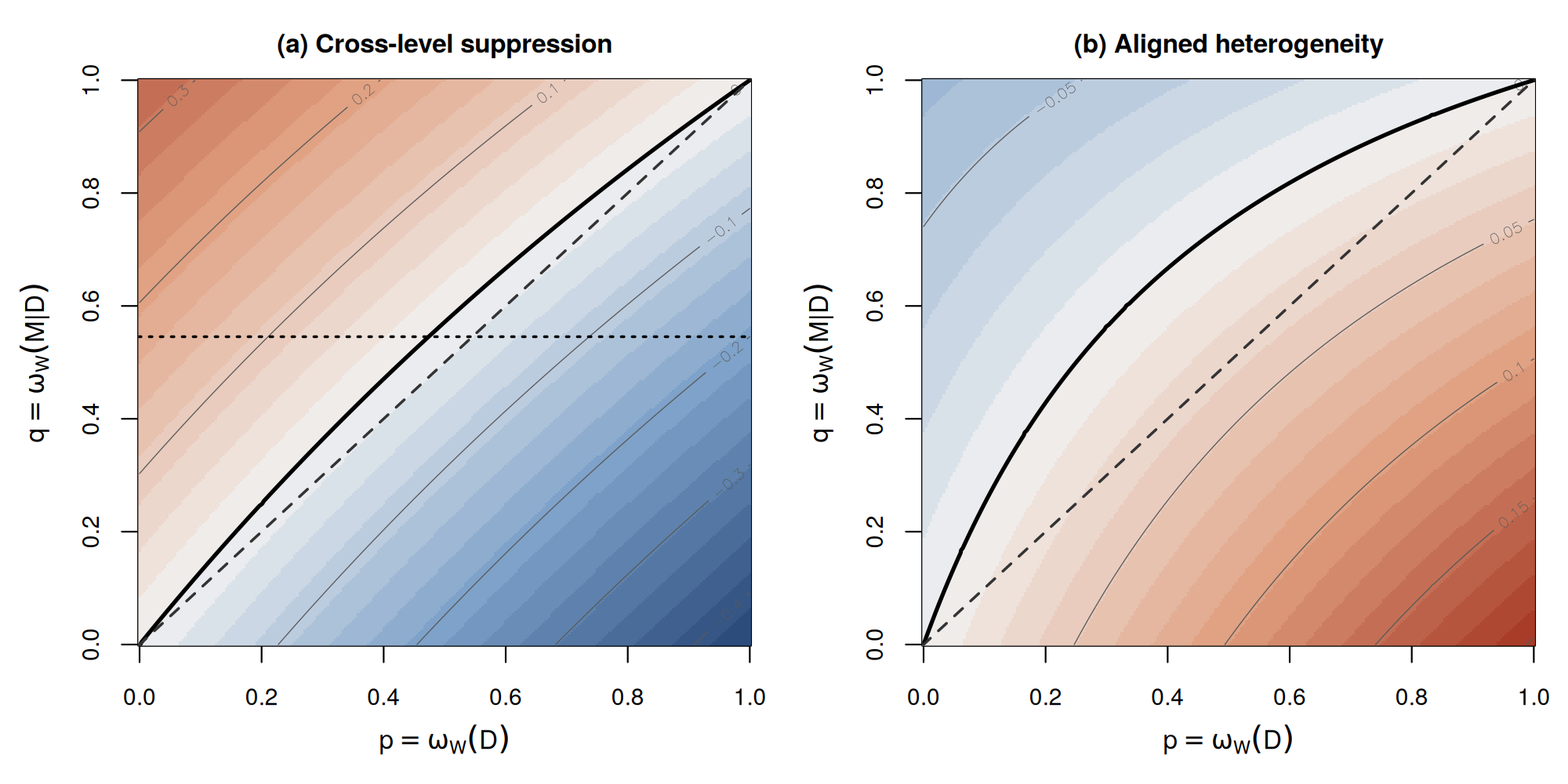}
\caption{Aggregation distortion \(A(p,q)\) over the design square. Shading and thin contours give the level of \(A\); the heavy black curve is the zero-distortion frontier \(\mathcal{Z}\); dotted lines mark sign reversals of the pooled \(a\)- and \(b\)-paths; the dashed diagonal is the equal-weight line \(p=q\) on which the textbook closed form is exact. Panel (a) uses the cross-level suppression configuration (\(\alpha_W=0.40\), \(\alpha_B=0.30\), \(\beta_{2W}=0.50\), \(\beta_{2B}=-0.60\)); panel (b) uses the aligned-heterogeneity configuration (\(\alpha_W=0.45\), \(\alpha_B=0.15\), \(\beta_{2W}=0.20\), \(\beta_{2B}=0.65\)).}
\label{fig:design-surface}
\end{figure}

Figure~\ref{fig:design-surface} shows the surface for two of the simulation configurations of Section 14, with contours, the zero-distortion frontier, the sign-reversal lines \(a(p)=0\) and \(b(q)=0\), and the equal-weight diagonal.

\section{8. Design Dependence Along the Equal-Weight
Diagonal}\label{design-dependence-along-the-equal-weight-diagonal}

It is also useful to examine the equal-weight case \(p=q\). Much of the
existing intuition corresponds to this diagonal, and the two-dimensional
surface then reduces to a one-dimensional curve.

\subsection{8.1 The Diagonal Slice}\label{the-diagonal-slice}

\textbf{Theorem 5 (Design Dependence Along the Diagonal).} \emph{On}
\(p = q = \omega\)\emph{, write}
\(g(\omega) = A(\omega,\omega) + T(\omega) = \left\lbrack \alpha_{B} + \omega\Delta_{\alpha} \right\rbrack\left\lbrack \beta_{2B} + \omega\Delta_{\beta} \right\rbrack\)\emph{.
Then:}

\emph{(a)} \(g\) \emph{is a quadratic in} \(\omega\) \emph{with leading
coefficient} \(\Delta_{\alpha}\Delta_{\beta}\)\emph{, and}
\(g(0) = IE_{B}\)\emph{,} \(g(1) = IE_{W}\)\emph{.}

\emph{(b) The stationary point is}
\(\omega_{v} = - \left( \alpha_{B}\Delta_{\beta} + \beta_{2B}\Delta_{\alpha} \right)/\left( 2\Delta_{\alpha}\Delta_{\beta} \right)\)
\emph{when} \(\Delta_{\alpha}\Delta_{\beta} \neq 0\)\emph{.}

\emph{(c)} \(g\) \emph{is monotone on} \(\lbrack 0,1\rbrack\) \emph{if
and only if} \(\omega_{v} \notin (0,1)\)\emph{. If}
\(\omega_{v} \in (0,1)\)\emph{, then} \(g\left( \omega_{v} \right)\)
\emph{lies strictly outside}
\(\left\lbrack \min\left( IE_{W},IE_{B} \right),\max\left( IE_{W},IE_{B} \right) \right\rbrack\)\emph{,
so the pooled estimand attains values that no level-specific indirect
effect attains.}

\emph{(d)} \(IE_{W}\) \emph{and} \(IE_{B}\) \emph{are invariant to}
\(\omega\)\emph{.}

\emph{\textbf{Proof.}} Appendix B. \(\blacksquare\)

Along the diagonal, both weights change together, so the bilinear
interaction appears as ordinary quadratic curvature. This behavior is
different from a single pooled regression coefficient, which must lie
between its within and between counterparts. A pooled indirect effect is
a product of two mixtures and can move outside that range. Section 14.2
gives an example with \(IE_{W}=0.090\) and \(IE_{B}=0.098\) where the
pooled value reaches \(0.128\) at an interior weight.

\textbf{Corollary 6 (Non-Comparability Across Designs).} \emph{Let two
studies of the same population share all structural paths but differ in
design, with weights}
\(\left( p^{(1)},q^{(1)} \right) \neq \left( p^{(2)},q^{(2)} \right)\)\emph{.
Then their pooled estimands differ whenever the gradient of Theorem 3 is
nonzero along a path connecting the two designs, while} \(IE_{W}\)
\emph{and} \(IE_{B}\) \emph{agree exactly.}

\emph{\textbf{Proof.}} Immediate from Theorem 3. \(\blacksquare\)

\textbf{Remark 9.} Corollary 6 matters when results are compared across
studies. If cluster-size distributions differ, pooled indirect effects
may target different quantities even when the structural paths are the
same. Some between-study heterogeneity can therefore be created by the
design itself. Level-specific effects avoid this problem, and Section 10
provides a scalar summary based on a fixed reference weight. This is
closely related to transportability of total effects (Degtiar \& Rose,
2023), except that here the changing quantity is partly determined by
the sampling design.

\subsection{8.2 A Local Fragility
Diagnostic}\label{a-local-fragility-diagnostic}

\textbf{Definition 6 (Design Sensitivity Index).} \emph{At the realized
within weight} \({\widehat{\omega}}_{W}\)\emph{,}

\[DSI = \frac{\left| g^{\prime}\left( {\widehat{\omega}}_{W} \right) \right|}{\left| IE_{W} \right| + \left| IE_{B} \right|},\quad\quad g^{\prime}(\omega) = \alpha_{B}\Delta_{\beta} + \beta_{2B}\Delta_{\alpha} + 2\omega\,\Delta_{\alpha}\Delta_{\beta}.\]

\(DSI\times0.01\) gives the proportional change in the pooled estimand
from a one-percentage-point shift of treatment variance from the between
to the within level, scaled by the total magnitude of the two
level-specific mechanisms. We do not propose a cutoff. \(DSI\) is a
descriptive measure, much like an ICC, and is best read together with
the region \(\mathcal{R}\) from Corollary 5 and the level-specific
intervals. Section 14.2 gives reference values for several simulation
settings.

When the two weights are allowed to move independently, the natural
generalization replaces the diagonal derivative with the gradient norm
of Theorem 3:

\[DSI_{2} = \frac{\parallel \nabla P\left( \widehat{p},\widehat{q} \right) \parallel}{\left| IE_{W} \right| + \left| IE_{B} \right|},\quad\quad\nabla P = \left( \Delta_{\alpha}\, b\left( \widehat{q} \right),\mspace{6mu}\Delta_{\beta}\, a\left( \widehat{p} \right) \right)^{\prime}.\]

\(DSI_{2}\) uses the steepest direction in the two-weight design space
rather than restricting attention to the diagonal. It is more
appropriate when \(\widehat p\) and \(\widehat q\) are well separated.
On the diagonal, it reduces to the same local idea as \(DSI\), up to the
choice of direction and scale.

\section{9. Cross-Level Suppression}\label{cross-level-suppression}

\textbf{Definition 7 (Cross-Level Suppression).} \emph{Cross-level
suppression occurs when}
\(sign\left( IE_{W} \right) \neq sign\left( IE_{B} \right)\)\emph{. A
pooled analysis exhibits complete suppression if}
\(plim\left( IE_{\text{pool}} \right) = 0\)\emph{, and partial
suppression if}
\(\left| plim\left( IE_{\text{pool}} \right) \right| < \min\left( \left| \omega_{W}IE_{W} \right|,\left| \omega_{B}IE_{B} \right| \right)\)\emph{.}

\textbf{Proposition 6 (Complete Suppression).} \(P(p,q) = 0\) \emph{if
and only if} \(a(p) = 0\) \emph{or} \(b(q) = 0\)\emph{. The zero sets
are the vertical line}
\(p^{*} = \alpha_{B}/\left( \alpha_{B} - \alpha_{W} \right)\) \emph{and
the horizontal line}
\(q^{*} = \beta_{2B}/\left( \beta_{2B} - \beta_{2W} \right)\)\emph{, and
each lies in} \((0,1)\) \emph{if and only if the corresponding pair of
level-specific coefficients has opposite signs. Complete suppression is
achievable by some design if and only if at least one path reverses sign
across levels.}

\emph{\textbf{Proof.}} Appendix B. \(\blacksquare\)

\textbf{Definition 8 (Sign-Reversal Frontier and Suppression Proximity
Index).} \emph{Restricting to the diagonal,}
\(\mathcal{F} = \{\omega \in (0,1):g(\omega) = 0\}\) \emph{and}
\(SPI = \min_{\omega \in \mathcal{F}}\left| {\widehat{\omega}}_{W} - \omega \right|\)\emph{.}
\(SPI \in \lbrack 0,1)\) \emph{is the distance from the realized design
to the nearest sign reversal. When}
\(\mathcal{F} = \varnothing\)\emph{,} \(SPI\) \emph{is undefined and the
pooled sign is design-invariant along the diagonal.}

\(DSI\) and \(SPI\) describe different features. \(DSI\) is local: it
measures how fast the pooled estimand changes at the observed design.
\(SPI\) measures how far the design is from a sign reversal. Both can be
small when the pooled effect is already close to zero.

\textbf{Remark 10.} \(SPI\) is similar in spirit to fragility thresholds
such as the Impact Threshold for a Confounding Variable (Frank, 2000).
The difference is that the perturbation here concerns observable,
design-driven variance weights rather than an unobserved confounder, so
the distance can be calculated directly from the fitted model.

\textbf{Remark 11 (Nonregularity).} \(SPI\) exists only when an
admissible sign-reversal frontier exists, and that event is estimated
from the data. An ordinary confidence interval is therefore not
appropriate. A more informative report gives the point value, when it
exists, together with the bootstrap frequency with which a frontier
exists. If no frontier exists, \(SPI\) is undefined. When plausible
ranges for the design weights are available, Corollary 5 provides the
more regular alternative.

\section{10. A Design-Standardized Indirect
Effect}\label{a-design-standardized-indirect-effect}

Because the pooled estimand changes with the design, a cross-study summary
should use a weight that is fixed independently of the realized sample.

\subsection{10.1 Definition}\label{definition}

\textbf{Definition 9 (Design-Standardized Indirect Effect).} \emph{For a
prespecified reference weight} \(r \in \lbrack 0,1\rbrack\)\emph{,}

\[IE_{\text{std}}(r) = r\, IE_{W} + (1 - r)\, IE_{B}.\]

\emph{The reference weight is fixed by the analyst in advance and is not
estimated from the realized sample. Natural choices include} \(r = 1/2\)
\emph{(equal emphasis on the two mechanisms), a value implied by a
target population's cluster-size distribution, a value fixed by
convention within a literature, or a policy-relevant value reflecting
the level at which an intervention would operate.}

The decomposed target from Section 5 is the special case
\(IE_{\text{dec}}=IE_{\text{std}}(p)\), where the reference weight is
taken from the realized design. Its design dependence therefore comes
from the choice of weight, not from the level-specific effects. Fixing
\(r\) in advance removes that source of variation.

\subsection{10.2 Invariance}\label{invariance}

\textbf{Theorem 6 (Design Standardization).} \emph{Suppose the
level-specific paths}
\(\left( \alpha_{W},\alpha_{B},\beta_{2W},\beta_{2B} \right)\) \emph{are
invariant across designs. Then for any fixed} \(r\)\emph{,}

\[\frac{\partial\, IE_{\text{std}}(r)}{\partial p} = \frac{\partial\, IE_{\text{std}}(r)}{\partial q} = 0,\]

\emph{whereas by Theorem 3 the pooled estimand satisfies}
\(\partial P/\partial p = \Delta_{\alpha}b(q)\) \emph{and}
\(\partial P/\partial q = \Delta_{\beta}a(p)\)\emph{, which vanish
identically only if} \(\Delta_{\alpha} = \Delta_{\beta} = 0\)\emph{.
Consequently two studies of the same population with different
cluster-size distributions report the same} \(IE_{\text{std}}(r)\)
\emph{and generally different pooled estimands.}

\emph{\textbf{Proof.}} Immediate from Definition 9 and Theorem 3.
\(\blacksquare\)

\textbf{Proposition 7 (Variance of the Standardized Effect).}
\emph{With}
\({\widehat{IE}}_{\text{std}}(r) = r{\widehat{IE}}_{W} + (1 - r){\widehat{IE}}_{B}\)\emph{,}

\[Var\left( {\widehat{IE}}_{\text{std}}(r) \right) = r^{2}V_{W} + (1 - r)^{2}V_{B} + 2r(1 - r)\, C_{WB},\]

\emph{where} \(C_{WB}\) \emph{is the asymptotic covariance of the two
level-specific estimators. Under balanced Gaussian designs}
\(C_{WB} = 0\)\emph{, because cluster means and within-cluster
deviations are independent; in general it need not vanish and should be
estimated by cluster bootstrap.}

\emph{\textbf{Proof.}} Appendix B. \(\blacksquare\)

A simulation check in Section 14.2 found little cross-level covariance in
the balanced design. Across 1,500 replications, the correlation between
the two level-specific estimators was \(-0.013\) (95\% CI
\(\lbrack-0.064,0.038\rbrack\)), and omitting \(C_{WB}\) changed the
standard error of \({\widehat{IE}}_{\text{std}}(r)\) by less than
0.5\% for \(r\in\{0.25,0.5,0.75\}\).

\subsection{10.3 Use}\label{use}

\textbf{Remark 12 (Choosing and reporting} \(r\)\textbf{).} The
reference weight should be chosen before examining the level-specific
estimates and reported with the result. Different values of \(r\)
define different standardized estimands, just as different reference
populations define different age-standardized rates. Because the
standardized effect is linear in \(r\), the full trajectory is easy to
show: it runs from \(IE_{B}\) at \(r=0\) to \(IE_{W}\) at \(r=1\), with
slope \(IE_{W}-IE_{B}\).

Standardization makes the summary comparable across designs; it does not
erase a real difference between within- and between-level mechanisms. If
\(IE_{W}\) and \(IE_{B}\) have opposite signs, both effects should be
reported rather than reduced to one scalar. \(MDI\) in Definition 11
provides useful context in that case.

\subsection{10.4 Reporting the Whole
Trajectory}\label{reporting-the-whole-trajectory}

Since
\(IE_{\text{std}}(r)=IE_{B}+r(IE_{W}-IE_{B})\), the full trajectory is
determined by the two endpoint effects. The slope
\(IE_{W}-IE_{B}\) shows how much the choice of reference weight matters.
A small slope means that reasonable values of \(r\) give similar
summaries. A large slope means that the reference weight has a
substantive influence and should be justified.

A simple companion measure is
\[
\frac{|IE_{W}-IE_{B}|}{|IE_{W}|+|IE_{B}|}.
\]
It is zero when the two level-specific effects agree and reaches one
when they are equal in magnitude and opposite in sign. In Section 15
the value is 0.0417, so the standardized summary changes little over the
reference-weight range. Proposition 7 gives a pointwise confidence
band. As before, standardization does not replace the level-specific
effects, and it does not strengthen their causal interpretation.

\section{11. Unification With Random-Coefficient
Heterogeneity}\label{unification-with-random-coefficient-heterogeneity}

\textbf{Proposition 8 (Pooled Limit Under Level Heterogeneity and Random
Paths).} \emph{Relax Definition 1 to permit random within-cluster
paths,} \(\alpha_{j} = \alpha_{W} + a_{j}\) \emph{and}
\(\beta_{2j} = \beta_{2W} + b_{j}\)\emph{. Assume: (i)}
\(\left( a_{j},b_{j} \right)\) \emph{has mean zero and is independent
of}
\(\left( {\bar{D}}_{j},D_{ij}^{W},\zeta_{j}^{M},\zeta_{j}^{Y},u_{ij},e_{ij} \right)\)\emph{;
(ii)} \(Cov\left( a_{j},b_{j} \right) = \tau_{ab}\) \emph{and}
\(Cov\left( b_{j},a_{j}^{2} \right) = 0\)\emph{, the latter holding
automatically under joint normality; (iii) the random paths have finite
second moments. Then the pooled path limits of Lemma 2 hold exactly, the
within-level indirect effect estimand becomes}

\[IE_{W}^{*} = \mathbb{E}\left\lbrack \alpha_{j}\beta_{2j} \right\rbrack = \alpha_{W}\beta_{2W} + \tau_{ab},\]

\emph{and total distortion relative to}
\(IE_{\text{dec}}^{*} = \omega_{W}(D)IE_{W}^{*} + \omega_{B}(D)IE_{B}\)
\emph{is}

\[A^{*} = \underset{\text{levels}}{\underbrace{- {Cov}_{\omega}\left( \alpha_{L},\beta_{L} \right)}}\mspace{6mu} + \mspace{6mu}\underset{\text{weight mismatch}}{\underbrace{\left( - \bar{\alpha}\Delta\Delta_{\beta} \right)}}\mspace{6mu} + \mspace{6mu}\underset{\text{clusters}}{\underbrace{\left( - \omega_{W}(D)\,\tau_{ab} \right)}}.\]

\emph{\textbf{Proof.}} Appendix B. \(\blacksquare\)

Condition (ii) removes one extra term. Without it, residualization can
make the random \(b\)-path depend on the squared random \(a\)-path,
leaving a remainder proportional to
\(Cov\left(b_{j},a_{j}^{2}\right)\).

The expression separates three sources of discrepancy. One comes from
mixing the within and between levels, one from using different weights
in the two pooled equations, and one from cluster-to-cluster covariance
in the random paths. Setting \(\tau_{ab}=0\) gives Theorem 2. Setting
\(\Delta_{\alpha}=\Delta_{\beta}=0\) and \(\Delta=0\) gives the
Kenny--Bauer case. A random-slope mediation model with the Kenny--Bauer
correction addresses the cluster-level covariance term, but not the
cross-level aggregation terms.

\section{12. The Diagnostic System}\label{the-diagnostic-system}

\textbf{Definition 10 (Aggregation Distortion and Percent Distortion).}

\[\widehat{AD} = {\widehat{IE}}_{\text{pool}} - \left( {\widehat{\omega}}_{W}{\widehat{IE}}_{W} + {\widehat{\omega}}_{B}{\widehat{IE}}_{B} \right),\quad\quad\widehat{PD} = \frac{\left| \widehat{AD} \right|}{\left| {\widehat{\omega}}_{W}{\widehat{IE}}_{W} + {\widehat{\omega}}_{B}{\widehat{IE}}_{B} \right|} \times 100\%.\]

\textbf{Definition 11 (Mechanism Dominance Index).}

\[MDI_{B} = \frac{\left| {\widehat{\omega}}_{B}{\widehat{IE}}_{B} \right|}{\left| {\widehat{\omega}}_{B}{\widehat{IE}}_{B} \right| + \left| {\widehat{\omega}}_{W}{\widehat{IE}}_{W} \right|},\quad\quad MDI_{W} = 1 - MDI_{B}.\]

\(PD\) can be unstable when the decomposed effect is close to zero. That
is also the setting in which suppression is most likely to attract
attention. For this reason, \(PD\) should be reported together with
\(\widehat{AD}\) and the intervals for the two level-specific indirect
effects.

\textbf{Reporting protocol.} A useful report contains the treatment and
mediator weights, the ICC on the same treatment scale, the weight
discrepancy, and the design coordinates
\((\widehat p,\widehat q)\). It should then give the within- and
between-level indirect effects, the pooled effect, and a
prespecified \(IE_{\text{std}}(r)\), preferably with cluster-bootstrap
intervals. For distortion, report \(\widehat{AD}\), \(\widehat{PD}\),
and the two components \({\widehat A}_{nl}\) and
\({\widehat A}_{wt}\). \(MDI\), \(DSI\), and the design region
\(\mathcal R\) add information about mechanism balance and design
fragility. \(SPI\) should be reported only when a sign-reversal frontier
exists, together with its bootstrap frontier-existence frequency. If the
two weights differ appreciably, the equal-weight formula is only a
descriptive approximation.

All of these quantities are functions of regression coefficients and
sample variances. The ancillary files contain the ECLS-K analysis code,
derived summaries used in the figures, and a base-R script that
reproduces all six figures. The values plotted for Studies 2 and 4 come
directly from Tables 4 and 6.

\section{13. Estimation and Asymptotic
Theory}\label{estimation-and-asymptotic-theory}

\subsection{13.1 The Residualized Person-Weighted Decomposition
Estimator}\label{the-residualized-person-weighted-decomposition-estimator}

\textbf{Definition 12 (Residualized Person-Weighted Decomposition
Estimator).} \emph{Let} \(\mathbf{C}_{ij}\) \emph{collect the covariates
and an intercept. The estimator proceeds in three ordered stages.}

\emph{Stage 1 (residualize, pooled across clusters). Form}

\[D_{ij}^{\bot} = D_{ij} - {\widehat{\mathbf{\Pi}}}_{D}^{\prime}\mathbf{C}_{ij},\quad\quad M_{ij}^{\bot} = M_{ij} - {\widehat{\mathbf{\Pi}}}_{M}^{\prime}\mathbf{C}_{ij},\quad\quad Y_{ij}^{\bot} = Y_{ij} - {\widehat{\mathbf{\Pi}}}_{Y}^{\prime}\mathbf{C}_{ij},\]

\emph{where each} \(\widehat{\mathbf{\Pi}}\) \emph{is the OLS
coefficient vector from the corresponding individual-level regression
on} \(\mathbf{C}_{ij}\)\emph{. Covariates are not decomposed at this
stage.}

\emph{Stage 2 (decompose the residualized variables). Form}

\[{\bar{D}}_{j}^{\bot} = \frac{1}{n_{j}}\sum_{i}^{}D_{ij}^{\bot},\quad\quad D_{ij,W}^{\bot} = D_{ij}^{\bot} - {\bar{D}}_{j}^{\bot},\]

\emph{and analogously for} \(M\) \emph{and} \(Y\)\emph{. The design
coordinate is}
\(\widehat{p} = \widehat{Var}\left( D_{ij,W}^{\bot} \right)/\widehat{Var}\left( D_{ij}^{\bot} \right)\)\emph{.}

\emph{Stage 3 (project, person-weighted). The} \(a\)\emph{-paths are the
projections}

\[{\widehat{\alpha}}_{W} = \frac{\sum_{ij}^{}D_{ij,W}^{\bot}M_{ij}^{\bot}}{\sum_{ij}^{}(D_{ij,W}^{\bot})^{2}},\quad\quad{\widehat{\alpha}}_{B} = \frac{\sum_{ij}^{}{\bar{D}}_{j}^{\bot}M_{ij}^{\bot}}{\sum_{ij}^{}({\bar{D}}_{j}^{\bot})^{2}},\]

\emph{both sums running over individuals, so that}
\({\widehat{\alpha}}_{B}\) \emph{is the cluster-means slope weighted by}
\(n_{j}\)\emph{. For the} \(b\)\emph{-paths, set}
\(M_{ij}^{\perp\!\!\!\perp} = M_{ij}^{\bot} - \widehat{\rho}\, D_{ij}^{\bot}\)
\emph{and}
\(Y_{ij}^{\perp\!\!\!\perp} = Y_{ij}^{\bot} - \widehat{\varsigma}\, D_{ij}^{\bot}\)\emph{,
decompose} \(M^{\perp\!\!\!\perp}\) \emph{as in Stage 2 to obtain}
\(\widehat{q}\)\emph{, and take}

\[{\widehat{\beta}}_{2W} = \frac{\sum_{ij}^{}M_{ij,W}^{\perp\!\!\!\perp}Y_{ij}^{\perp\!\!\!\perp}}{\sum_{ij}^{}(M_{ij,W}^{\perp\!\!\!\perp})^{2}},\quad\quad{\widehat{\beta}}_{2B} = \frac{\sum_{ij}^{}{\bar{M}}_{j}^{\perp\!\!\!\perp}Y_{ij}^{\perp\!\!\!\perp}}{\sum_{ij}^{}({\bar{M}}_{j}^{\perp\!\!\!\perp})^{2}}.\]

\emph{The level-specific indirect effect estimators are}
\({\widehat{IE}}_{W} = {\widehat{\alpha}}_{W}{\widehat{\beta}}_{2W}\)
\emph{and}
\({\widehat{IE}}_{B} = {\widehat{\alpha}}_{B}{\widehat{\beta}}_{2B}\)\emph{.}

\textbf{Remark 13 (Why the order matters).} Residualizing and
decomposing are not interchangeable steps, and the between-level
projection is person-weighted rather than an ordinary unweighted
regression on cluster means. With the ordering in Definition 12,
\[
\widehat{p}{\widehat{\alpha}}_{W}
 +(1-\widehat{p}){\widehat{\alpha}}_{B}
 ={\widehat{\alpha}}_{\text{pool}},
\]
with the analogous identity for the \(b\)-path. The common alternative
of decomposing first, adjusting within levels, and then using unweighted
cluster means does not satisfy these identities. Sections 14.5 and 15.1
show the resulting differences.

\subsection{13.2 Asymptotics}\label{asymptotics}

For inference on a product of coefficients, we use the usual delta method
(Sobel, 1982) applied to the cluster-level estimating equations.

\textbf{Theorem 7 (Consistency and Asymptotic Normality).} \emph{Under
(A1)--(A6) of Appendix A, as} \(J \rightarrow \infty\) \emph{with
cluster sizes bounded, the estimator is consistent for the
observed-projection estimands of Remark 3:}

\[{\widehat{IE}}_{W}\overset{p}{\rightarrow}IE_{W},\quad\quad{\widehat{IE}}_{B}\overset{p}{\rightarrow}IE_{B}.\]

\emph{Let}
\(\mathbf{\theta}_{W} = \left( \alpha_{W},\beta_{2W} \right)^{\prime}\)
\emph{and}
\(g\left( \mathbf{\theta} \right) = \theta_{1}\theta_{2}\)\emph{, and
define} \(\mathbf{\Sigma}_{W}\) \emph{by}

\[\sqrt{J}\,\left( {\widehat{\mathbf{\theta}}}_{W} - \mathbf{\theta}_{W} \right)\overset{d}{\rightarrow}\mathcal{N}\left( \mathbf{0},\mathbf{\Sigma}_{W} \right).\]

\emph{Then by the delta method}

\[\sqrt{J}\left( {\widehat{IE}}_{W} - IE_{W} \right)\overset{d}{\rightarrow}\mathcal{N}\left( 0,V_{W} \right),\quad\quad V_{W} = \nabla g\left( \mathbf{\theta}_{W} \right)^{\prime}\,\mathbf{\Sigma}_{W}\,\nabla g\left( \mathbf{\theta}_{W} \right),\]

\emph{with}
\(\nabla g\left( \mathbf{\theta}_{W} \right) = \left( \beta_{2W},\alpha_{W} \right)^{\prime}\)\emph{,
so that}
\(V_{W} = \beta_{2W}^{2}\Sigma_{W,11} + \alpha_{W}^{2}\Sigma_{W,22} + 2\alpha_{W}\beta_{2W}\Sigma_{W,12}\)\emph{.
The between-cluster result follows from the corresponding}
\(J\)\emph{-cluster estimating-equation system with each cluster's
contribution weighted by} \(n_{j}\)\emph{, equivalently from the
individual-level projection on repeated residualized cluster means,
with} \(\mathbf{\theta}_{B}\) \emph{and} \(\mathbf{\Sigma}_{B}\)
\emph{so defined. The} \(n_{j}\) \emph{weighting is not optional: it is
what Proposition 4 requires for the estimator to target the projection
that the pooled coefficient mixes.}

\emph{\textbf{Proof.}} Appendix B. \(\blacksquare\)

\textbf{Remark 14 (Scope of the consistency claim).} Theorem 7 is a
result for the observed-projection estimands. A latent between-level
target is a different problem, covered by Proposition 3 and Remark 4.
Bounded cluster sizes are enough here because the target itself is
defined using observed means; recovering a latent target would require
additional information or increasing cluster sizes.

\textbf{Remark 15 (The cross-equation covariance term).}
\({\widehat{\alpha}}_{W}\) and \({\widehat{\beta}}_{2W}\) are estimated
from different equations but use the same clusters and the same
mediator. Their sampling errors are therefore correlated, so
\(\Sigma_{W,12}\) should not be set to zero by default. In the
simulations of Section 14 the covariance is negative, and omitting it
makes the variance estimate conservative. In other settings the sign can
reverse. The term can be estimated from stacked estimating equations or
by cluster bootstrap. It is different from \(C_{WB}\), which links the
within- and between-level indirect-effect estimators.

\textbf{Remark 16 (Information asymmetry).} The within-level estimator
uses information from individual deviations, whereas the between-level
estimator is limited mainly by the number of clusters \(J\). Both are
\(\sqrt J\)-consistent under bounded cluster sizes, but their
finite-sample precision can be very different. Section 14.4 shows much
larger RMSE for \(IE_{B}\). Adding individuals within existing clusters
does not replace adding clusters when the target is between-level.

\section{14. Monte Carlo Evidence}\label{monte-carlo-evidence}

\subsection{14.1 Design}\label{design}

The simulations have four goals. Study 1 checks the decomposition in
Theorem 2 and the asymmetric homogeneity result in Corollary 2. Study 2
changes the mediator-side weight while holding the structural paths
fixed. Study 3 examines finite-cluster performance. A final numerical
check verifies the geometry in Section 7.

The data-generating process follows Definition 1 on the
\emph{observed} within--between decomposition. Treatment is first
generated from latent components, then converted to the observed cluster
mean and within-cluster deviation. The mediator and outcome are generated
from those observed quantities. This keeps the simulation target aligned
with the estimator. If the model were generated on latent components but
estimated from observed means, Proposition 3 would introduce a separate
measurement-error bias; Remark 17 shows how that pattern looks.

At \(n=20\), the variance components are chosen so that Proposition 1
gives \(\omega_{B}(D)=0.30\). The mediator equation includes nuisance
variance that makes \(\omega_{B}(M\mid D)\) differ from
\(\omega_{B}(D)\). Study 1 uses \(R=2{,}000\), \(J=50\), and \(n=20\).
Study 2 uses \(R=400\), \(J=60\), and \(n=25\). Study 3 uses \(R=600\)
with \(n_j\sim Uniform[10,30]\), drawn once for each cluster count and
then held fixed. Estimation follows Definition 12 with the
cluster-robust sandwich in Theorem 7. The accompanying base-R code
produces the reported results.

\subsection{14.2 Study 1: Verification of the
Decomposition}\label{study-1-verification-of-the-decomposition}

Configurations A--C and E vary the amount and direction of
within--between heterogeneity. Configuration F sets the \(a\)-paths
equal while allowing the \(b\)-paths to differ. Configuration G does the
reverse. F and G therefore isolate the two cases in Corollary 2.

\textbf{Table 1.} \emph{Verification of the two-component
decomposition.}

\begin{longtable}[]{@{}
  >{\raggedright\arraybackslash}p{(\columnwidth - 12\tabcolsep) * \real{0.1429}}
  >{\raggedright\arraybackslash}p{(\columnwidth - 12\tabcolsep) * \real{0.1429}}
  >{\raggedright\arraybackslash}p{(\columnwidth - 12\tabcolsep) * \real{0.1429}}
  >{\raggedright\arraybackslash}p{(\columnwidth - 12\tabcolsep) * \real{0.1429}}
  >{\raggedright\arraybackslash}p{(\columnwidth - 12\tabcolsep) * \real{0.1429}}
  >{\raggedright\arraybackslash}p{(\columnwidth - 12\tabcolsep) * \real{0.1429}}
  >{\raggedright\arraybackslash}p{(\columnwidth - 12\tabcolsep) * \real{0.1429}}@{}}
\toprule\noalign{}
\begin{minipage}[b]{\linewidth}\raggedright
Config.
\end{minipage} & \begin{minipage}[b]{\linewidth}\raggedright
Within, between \(a\)-path
\end{minipage} & \begin{minipage}[b]{\linewidth}\raggedright
Within, between \(b\)-path
\end{minipage} & \begin{minipage}[b]{\linewidth}\raggedright
Weight discrepancy
\end{minipage} & \begin{minipage}[b]{\linewidth}\raggedright
Nonlinearity component
\end{minipage} & \begin{minipage}[b]{\linewidth}\raggedright
Weight-mismatch component
\end{minipage} & \begin{minipage}[b]{\linewidth}\raggedright
Total distortion
\end{minipage} \\
\midrule\noalign{}
\endhead
\bottomrule\noalign{}
\endlastfoot
A & 0.35, 0.35 & 0.50, 0.50 & +0.0263 & -0.000000 & -0.000000 &
-0.000000 \\
B & 0.30, 0.50 & 0.40, 0.70 & +0.0322 & -0.012426 & +0.003467 &
-0.008959 \\
C & 0.40, 0.30 & 0.50, -0.60 & +0.0312 & -0.022737 & -0.012713 &
-0.035450 \\
E & 0.45, 0.15 & 0.20, 0.65 & +0.0328 & +0.027991 & +0.005345 &
+0.033336 \\
F & 0.35, 0.35 & 0.40, 0.70 & +0.0274 & +0.000000 & +0.002877 &
+0.002877 \\
G & 0.30, 0.50 & 0.50, 0.50 & +0.0336 & +0.000000 & -0.000000 &
+0.000000 \\
\end{longtable}

\emph{Note.} \(R = 2,000\), \(J = 50\), \(n = 20\). Components are
evaluated at mean empirical weights, which ranged over
\({\widehat{\omega}}_{B}(D) \in \lbrack 0.292,0.295\rbrack\) and
\({\widehat{\omega}}_{B}(M \mid D) \in \lbrack 0.321,0.326\rbrack\). In
every configuration \(A_{nl} + A_{wt}\) reproduced the direct difference
\(plim\left( IE_{\text{pool}} \right) - IE_{\text{dec}}\) to six decimal
places.

In every configuration, \(A_{nl}+A_{wt}\) matches the direct
calculation to numerical precision. F and G show the asymmetry clearly.
In F, equal \(a\)-paths make \(A_{nl}=0\), but the weight-mismatch term
still gives \(A=0.00288\). In G, equal \(b\)-paths make \(A=0\) even
though the \(a\)-paths differ by 0.20 and the two weights differ by
0.034.

\textbf{Table 2.} \emph{Level-specific estimator performance.}

\begin{longtable}[]{@{}
  >{\raggedright\arraybackslash}p{(\columnwidth - 12\tabcolsep) * \real{0.1429}}
  >{\raggedright\arraybackslash}p{(\columnwidth - 12\tabcolsep) * \real{0.1429}}
  >{\raggedright\arraybackslash}p{(\columnwidth - 12\tabcolsep) * \real{0.1429}}
  >{\raggedright\arraybackslash}p{(\columnwidth - 12\tabcolsep) * \real{0.1429}}
  >{\raggedright\arraybackslash}p{(\columnwidth - 12\tabcolsep) * \real{0.1429}}
  >{\raggedright\arraybackslash}p{(\columnwidth - 12\tabcolsep) * \real{0.1429}}
  >{\raggedright\arraybackslash}p{(\columnwidth - 12\tabcolsep) * \real{0.1429}}@{}}
\toprule\noalign{}
\begin{minipage}[b]{\linewidth}\raggedright
Config.
\end{minipage} & \begin{minipage}[b]{\linewidth}\raggedright
Within: bias
\end{minipage} & \begin{minipage}[b]{\linewidth}\raggedright
Within: RMSE
\end{minipage} & \begin{minipage}[b]{\linewidth}\raggedright
Within: coverage
\end{minipage} & \begin{minipage}[b]{\linewidth}\raggedright
Between: bias
\end{minipage} & \begin{minipage}[b]{\linewidth}\raggedright
Between: RMSE
\end{minipage} & \begin{minipage}[b]{\linewidth}\raggedright
Between: coverage
\end{minipage} \\
\midrule\noalign{}
\endhead
\bottomrule\noalign{}
\endlastfoot
A & +0.0002 & 0.0205 & 0.943 & -0.0009 & 0.0895 & 0.924 \\
B & -0.0002 & 0.0169 & 0.936 & +0.0016 & 0.1240 & 0.931 \\
C & +0.0005 & 0.0216 & 0.941 & -0.0019 & 0.0997 & 0.930 \\
E & -0.0003 & 0.0167 & 0.942 & +0.0014 & 0.1033 & 0.944 \\
F & +0.0002 & 0.0172 & 0.948 & +0.0000 & 0.1176 & 0.937 \\
G & -0.0005 & 0.0195 & 0.933 & +0.0011 & 0.0939 & 0.935 \\
\end{longtable}

\emph{Note.} Coverage is for nominal 95\% delta-method intervals
including the cross-equation covariance term of Remark 15.

\textbf{Table 3.} \emph{Distortion and design-sensitivity diagnostics.}

\begin{longtable}[]{@{}
  >{\raggedright\arraybackslash}p{(\columnwidth - 8\tabcolsep) * \real{0.2000}}
  >{\raggedright\arraybackslash}p{(\columnwidth - 8\tabcolsep) * \real{0.2000}}
  >{\raggedright\arraybackslash}p{(\columnwidth - 8\tabcolsep) * \real{0.2000}}
  >{\raggedright\arraybackslash}p{(\columnwidth - 8\tabcolsep) * \real{0.2000}}
  >{\raggedright\arraybackslash}p{(\columnwidth - 8\tabcolsep) * \real{0.2000}}@{}}
\toprule\noalign{}
\begin{minipage}[b]{\linewidth}\raggedright
Configuration
\end{minipage} & \begin{minipage}[b]{\linewidth}\raggedright
Mean percent distortion
\end{minipage} & \begin{minipage}[b]{\linewidth}\raggedright
Median percent distortion
\end{minipage} & \begin{minipage}[b]{\linewidth}\raggedright
Suppression detection rate
\end{minipage} & \begin{minipage}[b]{\linewidth}\raggedright
Design sensitivity index
\end{minipage} \\
\midrule\noalign{}
\endhead
\bottomrule\noalign{}
\endlastfoot
A & 1.8\% & 1.0\% & 1.4\% & 0.000 \\
B & 5.1\% & 4.0\% & 0.1\% & 0.436 \\
C & 60.4\% & 42.1\% & 97.5\% & 1.120 \\
E & 82.6\% & 31.6\% & 16.1\% & 0.337 \\
F & 5.6\% & 3.8\% & 1.1\% & 0.273 \\
G & 2.4\% & 1.4\% & 0.0\% & 0.250 \\
\end{longtable}

\emph{Note.} ``Suppression rate'' is the proportion of replications with
\(sign\left( {\widehat{IE}}_{W} \right) \neq sign\left( {\widehat{IE}}_{B} \right)\).
Mean \(PD\) exceeds median \(PD\) substantially in C and E because the
decomposed denominator approaches zero in some replications.

Bias is small for both level-specific estimators; the largest absolute
bias is 0.0019. Coverage for \({\widehat{IE}}_{W}\) ranges from 0.933 to
0.948. Coverage for \({\widehat{IE}}_{B}\) is a little lower, from 0.924
to 0.944, which is unsurprising with only \(J=50\) clusters.
Suppression is detected in 97.5\% of the replications under configuration
C and in at most 1.4\% of A, B, and G. Configuration E is less clear-cut
and produces a 16.1\% flag rate. The \(DSI\) values also separate the
designs in the expected direction.

Configuration G shows that zero distortion and design invariance are
different ideas. Because the \(b\)-paths are equal, pooled and
decomposed effects agree for every design. Yet \(DSI=0.250\), so that
common value still changes with the variance composition. Theorem 6
addresses this second problem by fixing the reference weight.

Configuration E provides the nonmonotone case from Theorem 5(c). Its
stationary point is \(\omega_v=0.472\), where the pooled estimand reaches
0.128, above both \(IE_W=0.090\) and \(IE_B=0.098\). Configuration C
provides the sign-reversal example from Definition 8:
\(\mathcal F=\{0.5455\}\) and \(SPI=0.160\) at the observed weight.

A separate check of Proposition 7 used 1,500 replications of
configuration B. The correlation between \({\widehat{IE}}_{W}\) and
\({\widehat{IE}}_{B}\) was \(-0.013\) (95\% CI
\(\lbrack-0.064,0.038\rbrack\), \(p=0.61\)). Dropping \(C_{WB}\) changed
the standard error of \({\widehat{IE}}_{\text{std}}(r)\) by at most
0.45\% for \(r\in\{0.25,0.5,0.75\}\).

\subsection{14.3 Study 2: Which Component
Dominates}\label{study-2-which-component-dominates}

Study 2 keeps the four structural paths at configuration B and changes
only \(\tau_M\), the cluster-level nuisance variance in the mediator
equation. Changing \(\tau_M\) changes the variance composition of the
residualized mediator and therefore changes \(\Delta\), but it leaves
the structural paths and \(\omega_B(D)\) fixed. In the design square,
the point moves vertically while \(p\) stays constant.

\textbf{Table 4.} \emph{Component dominance as the weight discrepancy
varies.}

\begin{longtable}[]{@{}
  >{\raggedright\arraybackslash}p{(\columnwidth - 12\tabcolsep) * \real{0.1429}}
  >{\raggedright\arraybackslash}p{(\columnwidth - 12\tabcolsep) * \real{0.1429}}
  >{\raggedright\arraybackslash}p{(\columnwidth - 12\tabcolsep) * \real{0.1429}}
  >{\raggedright\arraybackslash}p{(\columnwidth - 12\tabcolsep) * \real{0.1429}}
  >{\raggedright\arraybackslash}p{(\columnwidth - 12\tabcolsep) * \real{0.1429}}
  >{\raggedright\arraybackslash}p{(\columnwidth - 12\tabcolsep) * \real{0.1429}}
  >{\raggedright\arraybackslash}p{(\columnwidth - 12\tabcolsep) * \real{0.1429}}@{}}
\toprule\noalign{}
\begin{minipage}[b]{\linewidth}\raggedright
Mediator cluster SD
\end{minipage} & \begin{minipage}[b]{\linewidth}\raggedright
Between mediator weight
\end{minipage} & \begin{minipage}[b]{\linewidth}\raggedright
Weight discrepancy
\end{minipage} & \begin{minipage}[b]{\linewidth}\raggedright
Nonlinearity component
\end{minipage} & \begin{minipage}[b]{\linewidth}\raggedright
Weight-mismatch component
\end{minipage} & \begin{minipage}[b]{\linewidth}\raggedright
Total distortion
\end{minipage} & \begin{minipage}[b]{\linewidth}\raggedright
Weight-mismatch share
\end{minipage} \\
\midrule\noalign{}
\endhead
\bottomrule\noalign{}
\endlastfoot
0.05 & 0.049 & -0.240 & -0.01233 & -0.02573 & -0.03805 & 68\% \\
0.30 & 0.147 & -0.139 & -0.01226 & -0.01489 & -0.02714 & 55\% \\
0.55 & 0.318 & +0.031 & -0.01229 & +0.00332 & -0.00897 & 21\% \\
0.90 & 0.537 & +0.248 & -0.01233 & +0.02665 & +0.01432 & 68\% \\
1.40 & 0.731 & +0.441 & -0.01236 & +0.04739 & +0.03503 & 79\% \\
2.20 & 0.867 & +0.580 & -0.01228 & +0.06220 & +0.04992 & 84\% \\
\end{longtable}

\emph{Note.} \(R = 400\), \(J = 60\), \(n = 25\). Paths fixed at
\(\alpha_{W} = 0.30\), \(\alpha_{B} = 0.50\), \(\beta_{2W} = 0.40\),
\(\beta_{2B} = 0.70\);
\({\widehat{\omega}}_{B}(D) \in \lbrack 0.286,0.290\rbrack\) throughout.
``Wt. share'' is
\(\left| A_{wt} \right|/\left( \left| A_{nl} \right| + \left| A_{wt} \right| \right)\).

\(A_{nl}\) stays near \(-0.0123\), as expected, while \(A_{wt}\)
moves from \(-0.026\) to \(0.062\). The weight-mismatch term is the larger
component in five of the six settings. Total distortion also changes sign between \(\tau_M=0.55\) and \(0.90\), from
\(-0.0090\) to \(0.0143\), even though the structural paths and treatment
variance composition do not change. Geometrically, the design crosses
the zero-distortion frontier by moving in the \(q\) direction. The
equal-weight formula would give \(A=-0.0123\) for every row, so it misses
both the magnitude and, for half of the settings, the sign.

\begin{figure}[htbp]
\centering
\includegraphics[width=0.72\textwidth]{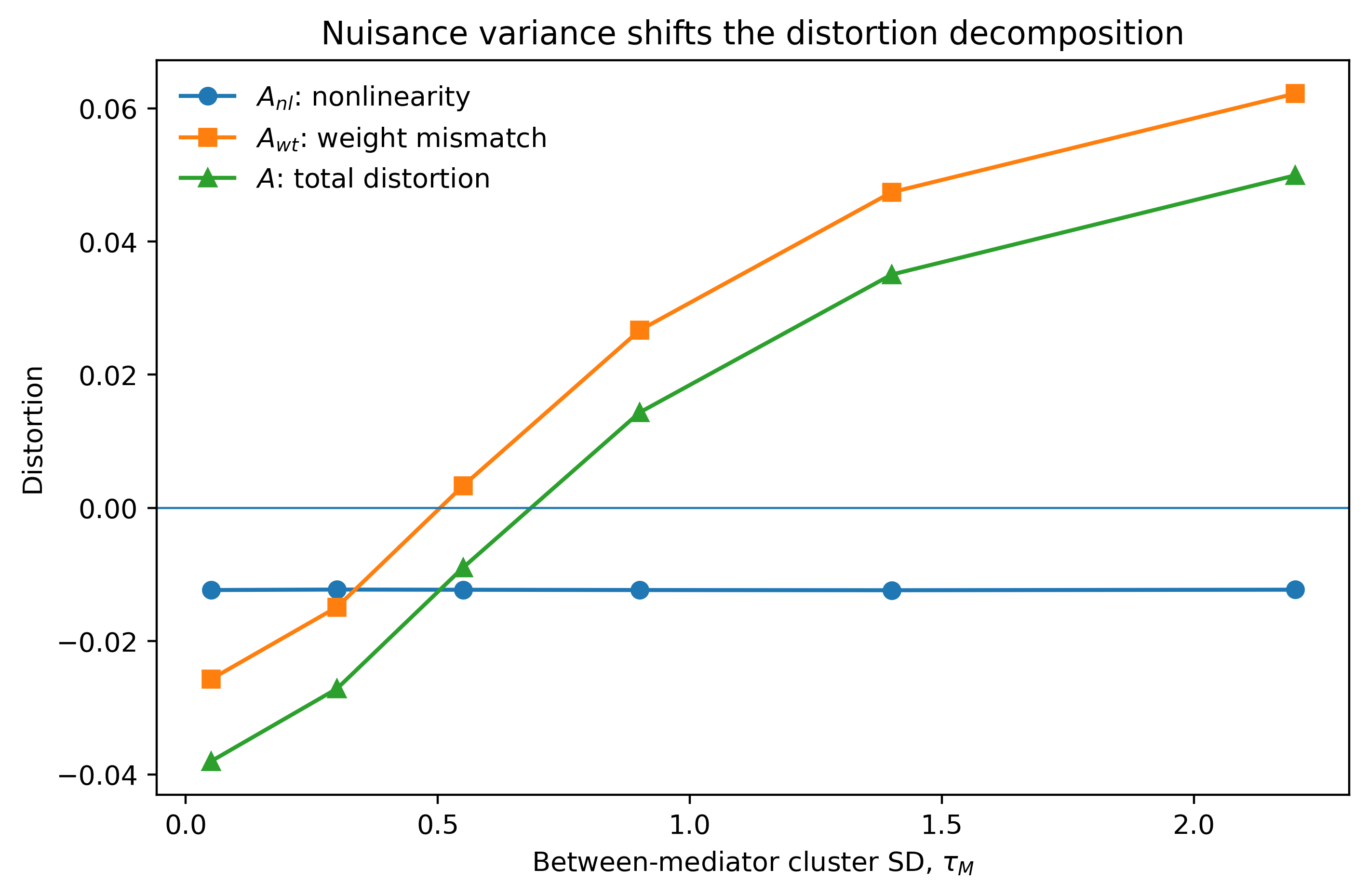}
\caption{Decomposition of aggregation distortion across the Study 2 nuisance-variance sweep. The nonlinearity component remains approximately constant, while the weight-mismatch component changes sign and grows, driving total distortion across zero between \(\tau_M=0.55\) and \(0.90\). Structural paths and the treatment variance composition are held fixed.}
\label{fig:component-dominance}
\end{figure}

Figure~\ref{fig:component-dominance} shows the same result graphically.
The nonlinearity component is nearly flat, while the weight-mismatch
component crosses zero and pulls the total distortion with it.

\subsection{14.4 Study 3: Finite-Cluster
Behavior}\label{study-3-finite-cluster-behavior}

\textbf{Table 5.} \emph{Level-specific estimator performance by cluster
count.}

\begin{longtable}[]{@{}
  >{\raggedright\arraybackslash}p{(\columnwidth - 12\tabcolsep) * \real{0.1429}}
  >{\raggedright\arraybackslash}p{(\columnwidth - 12\tabcolsep) * \real{0.1429}}
  >{\raggedright\arraybackslash}p{(\columnwidth - 12\tabcolsep) * \real{0.1429}}
  >{\raggedright\arraybackslash}p{(\columnwidth - 12\tabcolsep) * \real{0.1429}}
  >{\raggedright\arraybackslash}p{(\columnwidth - 12\tabcolsep) * \real{0.1429}}
  >{\raggedright\arraybackslash}p{(\columnwidth - 12\tabcolsep) * \real{0.1429}}
  >{\raggedright\arraybackslash}p{(\columnwidth - 12\tabcolsep) * \real{0.1429}}@{}}
\toprule\noalign{}
\begin{minipage}[b]{\linewidth}\raggedright
Number of clusters
\end{minipage} & \begin{minipage}[b]{\linewidth}\raggedright
Between: bias
\end{minipage} & \begin{minipage}[b]{\linewidth}\raggedright
Between: RMSE
\end{minipage} & \begin{minipage}[b]{\linewidth}\raggedright
Between: coverage
\end{minipage} & \begin{minipage}[b]{\linewidth}\raggedright
Within: bias
\end{minipage} & \begin{minipage}[b]{\linewidth}\raggedright
Within: RMSE
\end{minipage} & \begin{minipage}[b]{\linewidth}\raggedright
Within: coverage
\end{minipage} \\
\midrule\noalign{}
\endhead
\bottomrule\noalign{}
\endlastfoot
15 & -0.0166 & 0.2508 & 0.883 & +0.0007 & 0.0293 & 0.932 \\
25 & -0.0107 & 0.1765 & 0.918 & -0.0016 & 0.0218 & 0.952 \\
40 & -0.0020 & 0.1385 & 0.927 & +0.0005 & 0.0181 & 0.937 \\
60 & +0.0033 & 0.1157 & 0.925 & -0.0000 & 0.0158 & 0.937 \\
100 & -0.0052 & 0.0886 & 0.928 & -0.0005 & 0.0116 & 0.950 \\
\end{longtable}

\emph{Note.} \(R = 600\); \(n_{j} \sim Uniform\lbrack 10,30\rbrack\),
drawn once per cluster count and held fixed across replications;
configuration B paths.

As \(J\) increases from 15 to 100, RMSE for
\({\widehat{IE}}_{B}\) falls by a factor of 2.8. The within-level
estimator is much more precise throughout. Bias in
\({\widehat{IE}}_{B}\) is \(-0.0166\) at \(J=15\) and is close to zero
by about \(J=40\), consistent with finite-sample bias in a product of
estimated coefficients. Coverage improves from 0.883 to about 0.925 but
remains somewhat below 0.95. This appears to be a small-\(J\) issue for
the sandwich variance rather than an inconsistency in the estimator. A
CR2-type correction (Pustejovsky \& Tipton, 2018) may help, but it is not
studied here. For small \(J\), the cluster bootstrap is the safer basis
for between-level inference, especially when suppression depends on the
sign of \({\widehat{IE}}_{B}\).

\textbf{Remark 17 (A useful diagnostic).} In an earlier simulation
version, the model was generated on latent cluster components but
estimated with observed cluster means. The between-level bias was about
\(-0.04\) and did not shrink as \(J\) increased. Its magnitude closely
matched the reliability factor \(\lambda_n=0.877\) in Proposition 3.
That pattern is a useful warning: bias that persists as the number of
clusters grows can indicate a latent-versus-observed measurement-error
mismatch rather than a small-sample variance problem.

\subsection{14.5 Study 4: Construction-Induced Spurious
Distortion}\label{study-4-construction-induced-spurious-distortion}

Section 15.1 shows a noticeable difference between the conventional and
identity-consistent constructions in the ECLS-K example. Study 4 checks
whether the same pattern appears systematically in simulation and how it
changes with cluster-size imbalance and covariate--treatment
association.

The design crosses four cluster-size distributions with
\(\rho_X\in\{0,0.3,0.6\}\), using \(R=200\) replications per cell. Each
replication is analyzed twice. The identity-consistent version
residualizes first, then decomposes, and uses person-weighted
between-level projections. The conventional version decomposes first,
adjusts for the covariate separately within levels, computes weights from
the raw treatment, and uses unweighted cluster-means regressions. The
underlying structural paths are the same in both analyses.

\textbf{Table 6.} \emph{Construction-induced spurious distortion.}

\begin{longtable}[]{@{}
  >{\raggedright\arraybackslash}p{(\columnwidth - 14\tabcolsep) * \real{0.1250}}
  >{\raggedright\arraybackslash}p{(\columnwidth - 14\tabcolsep) * \real{0.1250}}
  >{\raggedright\arraybackslash}p{(\columnwidth - 14\tabcolsep) * \real{0.1250}}
  >{\raggedright\arraybackslash}p{(\columnwidth - 14\tabcolsep) * \real{0.1250}}
  >{\raggedright\arraybackslash}p{(\columnwidth - 14\tabcolsep) * \real{0.1250}}
  >{\raggedright\arraybackslash}p{(\columnwidth - 14\tabcolsep) * \real{0.1250}}
  >{\raggedright\arraybackslash}p{(\columnwidth - 14\tabcolsep) * \real{0.1250}}
  >{\raggedright\arraybackslash}p{(\columnwidth - 14\tabcolsep) * \real{0.1250}}@{}}
\toprule\noalign{}
\begin{minipage}[b]{\linewidth}\raggedright
Cluster sizes
\end{minipage} & \begin{minipage}[b]{\linewidth}\raggedright
\(CV\left( n_{j} \right)\)
\end{minipage} & \begin{minipage}[b]{\linewidth}\raggedright
\(\rho_{X}\)
\end{minipage} & \begin{minipage}[b]{\linewidth}\raggedright
Consistent \(|A|\)
\end{minipage} & \begin{minipage}[b]{\linewidth}\raggedright
Conv. \(|A|\)
\end{minipage} & \begin{minipage}[b]{\linewidth}\raggedright
Ratio
\end{minipage} & \begin{minipage}[b]{\linewidth}\raggedright
Conv. identity error
\end{minipage} & \begin{minipage}[b]{\linewidth}\raggedright
Consistent identity error
\end{minipage} \\
\midrule\noalign{}
\endhead
\bottomrule\noalign{}
\endlastfoot
Balanced & 0.00 & 0.0 & 0.000456 & 0.000477 & 1.05 & 0.55\% & 6.3e-18 \\
Balanced & 0.00 & 0.3 & 0.000497 & 0.000662 & 1.33 & 8.44\% & 6.3e-18 \\
Balanced & 0.00 & 0.6 & 0.000669 & 0.001074 & 1.60 & 20.45\% &
8.0e-18 \\
Mild & 0.28 & 0.0 & 0.000446 & 0.000619 & 1.39 & 4.43\% & 5.6e-18 \\
Mild & 0.28 & 0.3 & 0.000482 & 0.000757 & 1.57 & 9.79\% & 7.6e-18 \\
Mild & 0.28 & 0.6 & 0.000669 & 0.001130 & 1.69 & 21.62\% & 7.6e-18 \\
Moderate & 0.87 & 0.0 & 0.000495 & 0.001089 & 2.20 & 13.09\% &
7.9e-18 \\
Moderate & 0.87 & 0.3 & 0.000502 & 0.001209 & 2.41 & 16.24\% &
8.0e-18 \\
Moderate & 0.87 & 0.6 & 0.000612 & 0.001499 & 2.45 & 25.23\% &
8.6e-18 \\
Extreme & 2.16 & 0.0 & 0.000862 & 0.001767 & 2.05 & 37.73\% & 6.1e-18 \\
Extreme & 2.16 & 0.3 & 0.000856 & 0.001798 & 2.10 & 58.96\% & 7.8e-18 \\
Extreme & 2.16 & 0.6 & 0.000953 & 0.002004 & 2.10 & 125.88\% &
7.3e-18 \\
\end{longtable}

Note. The final three rows use a deliberately extreme cluster-size
distribution with \(CV(n_j)=2.16\) as a stress test; they are not
calibrated to the ECLS-K application. Three quantities should not be
conflated. Identity error is a discrepancy in a single pooled path,
expressed as a percentage of the fitted coefficient. Aggregation
distortion is a discrepancy in the indirect effect, in effect-size
units. The ratio is conventional absolute distortion divided by
identity-consistent absolute distortion, a unitless inflation factor.
Structural paths are held fixed across all rows.

The identity-consistent construction satisfies Lemma 2 to machine
precision in every cell. The conventional construction does not. Its
identity error is only 0.55\% in the balanced, covariate-orthogonal case,
but it rises as covariate association and cluster-size imbalance enter
the design. For example, it reaches 20.45\% under balance with
\(\rho_X=0.6\), 37.73\% under severe imbalance with \(\rho_X=0\), and
125.88\% when both features are extreme.

In this simulation grid, the conventional construction also gives a
larger absolute distortion in every cell. The ratio ranges from 1.05 to
2.45. The increase is not monotone in cluster-size imbalance: the ratio
peaks at the moderate-imbalance setting and falls somewhat in the most
extreme rows. The main point is therefore not monotonicity. It is that
once the paths and weights are constructed on different scales, the
reported distortion contains variation that is not part of the
aggregation identity. The identity check in Section 16.1 detects this
before the distortion is interpreted.

\begin{figure}[htbp]
\centering
\includegraphics[width=0.76\textwidth]{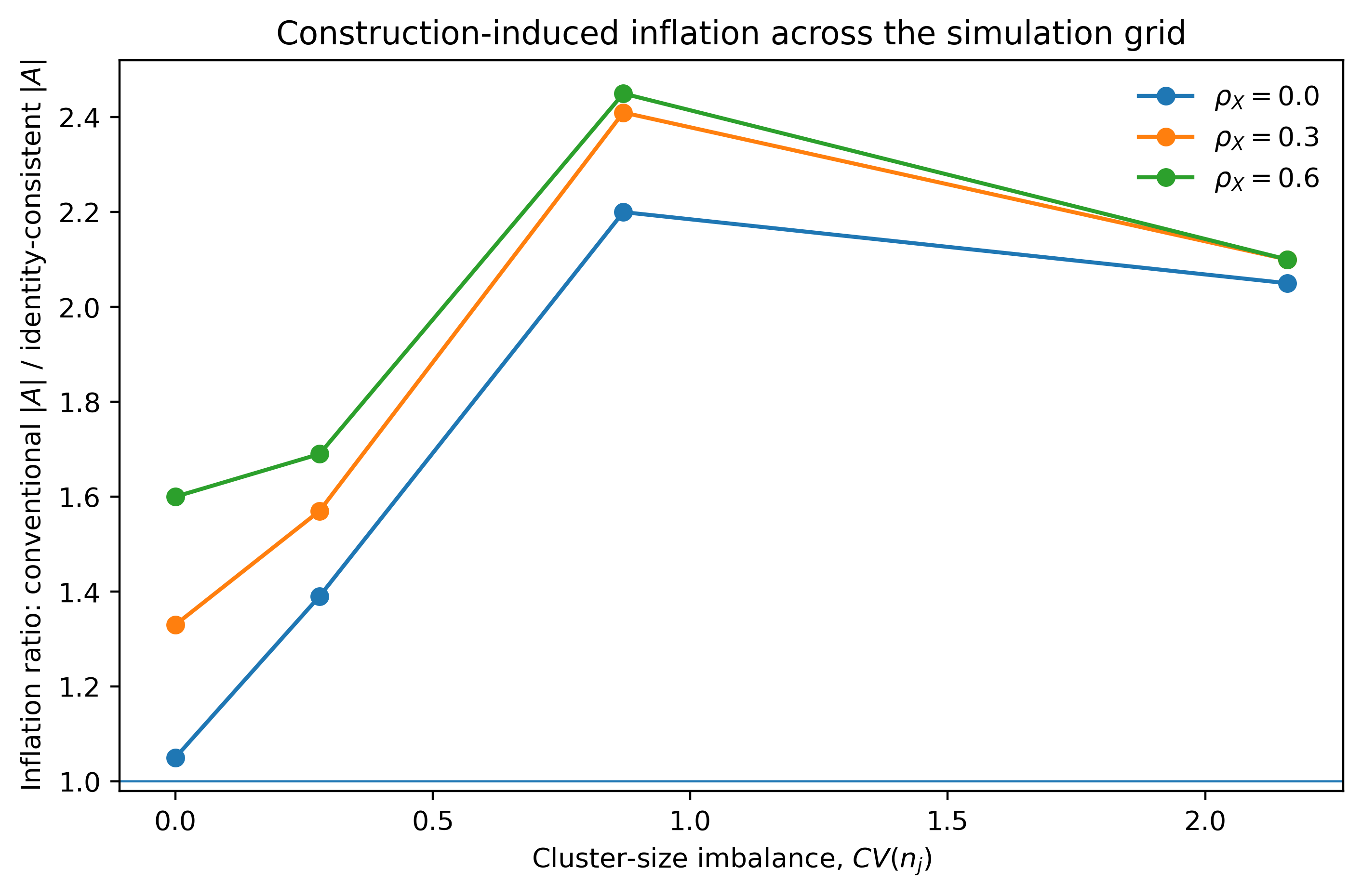}
\caption{Construction-induced inflation across the Study 4 simulation grid. The vertical axis is conventional absolute distortion divided by identity-consistent absolute distortion. Each line corresponds to a level of covariate--treatment correlation \(\rho_X\); the horizontal axis is the coefficient of variation of cluster size.}
\label{fig:construction-inflation}
\end{figure}

Figure~\ref{fig:construction-inflation} plots the same ratios. The
balanced, covariate-orthogonal case is close to one. The ratios are much
larger once covariate association or cluster-size imbalance is present,
although the change with imbalance is not monotone.

\subsection{14.6 Verification of the
Geometry}\label{verification-of-the-geometry}

Because the results in Section 7 are algebraic, they were checked
numerically rather than by Monte Carlo sampling. For configuration C,
the numerical second derivatives of \(P\) were
\(\partial^{2}P/\partial p^{2} = - 1.4 \times 10^{- 7}\) and
\(\partial^{2}P/\partial q^{2} = 2.1 \times 10^{- 7}\) against a
theoretical zero, while the cross-partial was \(0.110000\) against
\(\Delta_{\alpha}\Delta_{\beta} = 0.110000\). The four corner values
were \(A(0,0) = 0\), \(A(1,1) = 5.6 \times 10^{- 17}\),
\(A(0,1) = + 0.3300 = \alpha_{B}\Delta_{\beta}\), and
\(A(1,0) = - 0.4400 = - \alpha_{W}\Delta_{\beta}\). A \(401 \times 401\)
grid search gave \(\sup|A| = 0.440000\), matching the closed-form bound
\(\left| \Delta_{\beta} \right|\max\left( \left| \alpha_{W} \right|,\left| \alpha_{B} \right| \right) = 0.440000\)
to six decimals. Residual distortion along the frontier \(q_{0}(p)\) of
Corollary 4 was at most \(2.8 \times 10^{- 17}\).

\subsection{14.7 Summary}\label{summary}

Taken together, the simulations support the main algebraic results.
The decomposition is exact, the two homogeneity conditions behave
differently, and the weight-mismatch term can be the main source of
distortion. Changing a nuisance variance can even reverse the sign of
the total distortion while the structural paths remain fixed. The
bilinear geometry and corner bounds match their closed forms. The
simulation also shows why the construction order matters: the
identity-consistent residual stays at machine precision, whereas the
conventional construction can create substantial extra discrepancy.
Finally, the level-specific estimators are nearly unbiased for their
observed-projection targets, but the between-level estimator is much
less precise.

\section{15. Numerical Illustration}\label{numerical-illustration}

We next apply the diagnostics to the Early Childhood Longitudinal Study,
Kindergarten Class of 2010--11 (National Center for Education Statistics [NCES], 2019). The purpose is to show how
the unequal-weight problem looks in a real multilevel data set. This is
not a substantive causal analysis. Assumption 1 is not established by
the observational design, the analysis uses complete cases, and the NCES
complex survey weights are not incorporated.

The example asks whether parent-reported approaches to learning mediate
the association between family socioeconomic status and later
mathematics achievement, adjusting for baseline mathematics, child sex,
and race/ethnicity. Clusters are defined by \texttt{S2\_ID}, the
spring-kindergarten school identifier. NCES 9000-series status codes are
treated as non-school values. Thus, the within- and between-school
quantities refer to children who shared a baseline school; they need not
have attended the same school at later waves. There are 8,699 complete
cases in 937 valid baseline schools. Requiring at least five complete
children per school leaves 8,363 children in 753 schools. The harmonic
mean cluster size is 9.89, the maximum is 21, and \(CV(n_j)=0.321\).
Uncertainty is assessed with 1,000 school-cluster bootstrap
replications.

\textbf{Table 7.} \emph{Diagnostic protocol applied to ECLS-K:2011.}

\begin{longtable}[]{@{}
  >{\raggedright\arraybackslash}p{(\columnwidth - 4\tabcolsep) * \real{0.3333}}
  >{\raggedright\arraybackslash}p{(\columnwidth - 4\tabcolsep) * \real{0.3333}}
  >{\raggedright\arraybackslash}p{(\columnwidth - 4\tabcolsep) * \real{0.3333}}@{}}
\toprule\noalign{}
\begin{minipage}[b]{\linewidth}\raggedright
Symbol
\end{minipage} & \begin{minipage}[b]{\linewidth}\raggedright
Quantity
\end{minipage} & \begin{minipage}[b]{\linewidth}\raggedright
Estimate {[}95\% bootstrap CI{]}
\end{minipage} \\
\midrule\noalign{}
\endhead
\bottomrule\noalign{}
\endlastfoot
& Residualized-treatment variance structure & \\
\({\widehat{\omega}}_{B}(D)\) & Between-cluster weight, residualized SES
& 0.277 {[}0.256, 0.296{]} \\
\(\widehat{ICC}(D)\) & Intraclass correlation, residualized SES &
0.206 \\
\(\delta\) & Weight-minus-ICC gap (Prop. 1) & 0.071 \\
& Identity-consistent design coordinates & \\
\(\widehat{p}\) & Within weight, residualized treatment & 0.723
{[}0.704, 0.744{]} \\
\(\widehat{q}\) & Within weight, residualized mediator & 0.896 {[}0.887,
0.906{]} \\
\(\widehat{\Delta}\) & Weight discrepancy (p - q) & -0.173 {[}-0.193,
-0.151{]} \\
& Level-specific paths (person-weighted) & \\
\({\widehat{\alpha}}_{W},\ {\widehat{\beta}}_{2W}\) & Within
baseline-school paths & 0.0998 {[}0.0673, 0.1296{]}; 0.0903 {[}0.0730,
0.1071{]} \\
\({\widehat{\alpha}}_{B},\ {\widehat{\beta}}_{2B}\) & Between
baseline-school paths & 0.0707 {[}0.0228, 0.1181{]}; 0.1385 {[}0.0697,
0.2073{]} \\
& b-path difference (between - within) & +0.0482 {[}-0.0252,
0.1202{]} \\
& Estimands & \\
\({\widehat{IE}}_{W}\) & Within-baseline-school indirect effect &
0.00901 {[}0.00575, 0.01233{]} \\
\({\widehat{IE}}_{B}\) & Between-baseline-school indirect effect &
0.00980 {[}0.00283, 0.01970{]} \\
\({\widehat{IE}}_{\text{dec}}\) & Decomposed target at realized design &
0.00923 {[}0.00609, 0.01265{]} \\
\({\widehat{IE}}_{\text{pool}}\) & Pooled indirect effect & 0.00874
{[}0.00584, 0.01174{]} \\
\(IE_{\text{std}}(0.5)\) & Design-standardized effect (r = 0.50) &
0.00940 {[}0.00531, 0.01443{]} \\
& Distortion & \\
\(\widehat{AD}\) & Aggregation distortion & -0.000485 {[}-0.001959,
0.000343{]} \\
\(A_{nl}\) & Nonlinearity component & +0.000281 {[}-0.000360,
0.001275{]} \\
\(A_{wt}\) & Weight-mismatch component & -0.000766 {[}-0.001973,
0.000395{]} \\
\(\widehat{PD}\) & Percent distortion & 5.26\% (descriptive; CI not
reported) \\
& Geometry & \\
\(\pm \left| \Delta_{\alpha}\Delta_{\beta} \right|\) & Hessian
eigenvalues (Thm. 3) & +/-0.001402 \\
\(DSI_{2}\) & Design sensitivity, gradient norm & 0.278 {[}0.094,
0.750{]} \\
\(\mathcal{R}_{P}\) & Sharp bounds on pooled effect & {[}0.00804,
0.00948{]} \\
\(\mathcal{R}_{A}\) & Sharp bounds on distortion & {[}-0.001269,
0.000327{]} \\
\end{longtable}

Note. N = 8,363 children in J = 753 spring-kindergarten baseline schools
after requiring at least five complete children per school; harmonic
mean cluster size = 9.89, maximum cluster size = 21, and
\(CV(n_j)=0.321\). Percentile 95\% confidence intervals are based on 1,000
school-cluster bootstrap replications of the identity-consistent
residualize-first, person-weighted estimator. Focal continuous variables
are standardized. The ICC row is descriptive; bracketed intervals
elsewhere are bootstrap percentile intervals unless noted. The treatment
weight and ICC are reported on the same covariate-residualized treatment
scale. PD is reported descriptively without a bootstrap interval because
the absolute-value transformation folds the AD distribution at zero.
Sharp bounds evaluate the four corners of the stated design region and
are not bootstrap intervals.

Table 7 gives the identity-consistent estimates and bootstrap intervals.
For residualized treatment, the within weight is \(p=0.723\) (95\% CI
{[}0.704, 0.744{]}). For the residualized mediator, it is \(q=0.896\)
(95\% CI {[}0.887, 0.906{]}). Their difference is
\(\Delta=p-q=-0.173\) (95\% CI {[}-0.193, -0.151{]}), so the two
weighting systems are clearly different. The between-school \(b\)-path
is 0.139 at the point estimate, compared with 0.090 within schools, but
the bootstrap interval for their difference {[}-0.0252, 0.1202{]}
includes zero. The application therefore gives strong evidence of
unequal weighting, but not of a population difference in the two
\(b\)-paths.

Several numbers connect the example to the theory. On the
covariate-residualized treatment, the between-cluster weight is 0.277
while the ICC is 0.206. The difference of 0.071 is large enough that the
two should not be treated as interchangeable. The pooled indirect effect
is 0.00874 and the decomposed target is 0.00923, giving
\(AD=-0.000485\) and \(PD=5.26\%\). At the point estimate, the
nonlinearity component is \(+0.000281\) and the weight-mismatch component
is \(-0.000766\), so they partly cancel. The Hessian eigenvalues are
\(\pm0.001402\), as predicted by the saddle geometry. Over the stated
design rectangle, the pooled effect remains positive
({[}0.00804, 0.00948{]}), but the distortion range crosses zero
({[}-0.00127, 0.000327{]}). The sign of the mediated association is
therefore stable over that region even though the sign of the
aggregation distortion is not.

\subsection{15.1 Two Between-Level Estimands, and a Cautionary
Result}\label{two-between-level-estimands-and-a-cautionary-result}

Proposition 4 and Remark 5 require the decomposition to be built on the
same residualized scale as the pooled regression: residualize first,
decompose second, and use person-weighted between-level projections. A
common alternative adjusts covariates separately within levels and fits
an unweighted regression to the school means. That alternative can be a
reasonable descriptive analysis, but its coefficients are not the
between-level coefficients mixed by the pooled estimator.

\textbf{Table 8.} \emph{Person-weighted and cluster-weighted
between-level estimands.}

\begin{longtable}[]{@{}
  >{\raggedright\arraybackslash}p{(\columnwidth - 6\tabcolsep) * \real{0.2500}}
  >{\raggedright\arraybackslash}p{(\columnwidth - 6\tabcolsep) * \real{0.2500}}
  >{\raggedright\arraybackslash}p{(\columnwidth - 6\tabcolsep) * \real{0.2500}}
  >{\raggedright\arraybackslash}p{(\columnwidth - 6\tabcolsep) * \real{0.2500}}@{}}
\toprule\noalign{}
\begin{minipage}[b]{\linewidth}\raggedright
Quantity
\end{minipage} & \begin{minipage}[b]{\linewidth}\raggedright
Person-weighted
\end{minipage} & \begin{minipage}[b]{\linewidth}\raggedright
Cluster-weighted
\end{minipage} & \begin{minipage}[b]{\linewidth}\raggedright
Difference
\end{minipage} \\
\midrule\noalign{}
\endhead
\bottomrule\noalign{}
\endlastfoot
a path & 0.07071 & 0.08327 & -0.01256 \\
b path & 0.13854 & 0.11942 & +0.01911 \\
Indirect effect & 0.00980 & 0.00994 & -0.00015 \\
\end{longtable}

Note. Both columns residualize first and then decompose; they differ
only in whether the between-level regression is person-weighted or gives
each school equal weight. The person- and cluster-weighted between
indirect effects are 0.00980 and 0.00994, a difference of -0.00015 (95\%
bootstrap CI {[}-0.00375, 0.00331{]}). Informative cluster size is
therefore not consequential for the indirect-effect product in this
application, although Proposition 4 still requires person weighting for
the exact pooled identity.

\textbf{Table 9.} \emph{Verification of the Mundlak identity.}

\begin{longtable}[]{@{}
  >{\raggedright\arraybackslash}p{(\columnwidth - 6\tabcolsep) * \real{0.2500}}
  >{\raggedright\arraybackslash}p{(\columnwidth - 6\tabcolsep) * \real{0.2500}}
  >{\raggedright\arraybackslash}p{(\columnwidth - 6\tabcolsep) * \real{0.2500}}
  >{\raggedright\arraybackslash}p{(\columnwidth - 6\tabcolsep) * \real{0.2500}}@{}}
\toprule\noalign{}
\begin{minipage}[b]{\linewidth}\raggedright
Path
\end{minipage} & \begin{minipage}[b]{\linewidth}\raggedright
Fitted pooled
\end{minipage} & \begin{minipage}[b]{\linewidth}\raggedright
Identity, person-weighted
\end{minipage} & \begin{minipage}[b]{\linewidth}\raggedright
Identity, cluster-weighted
\end{minipage} \\
\midrule\noalign{}
\endhead
\bottomrule\noalign{}
\endlastfoot
a path & 0.091734 & 0.091734 (error -1.39e-17) & 0.095211 (+3.79\%) \\
b path & 0.095316 & 0.095316 (error 0) & 0.093333 (-2.08\%) \\
\end{longtable}

Note. Under the person-weighted construction both pooled-path identities
hold to machine precision. Substituting cluster-weighted between paths
gives 0.095211 for the a-path and 0.093333 for the b-path, corresponding
to identity errors of +3.79\% and -2.08\% relative to the fitted pooled
coefficients.

Table 10 compares the two constructions while holding the fitted pooled
indirect effect fixed at 0.00874. The conventional decomposition gives a
target of 0.00960 and \(AD=-0.000854\). The identity-consistent
construction gives 0.00923 and \(AD=-0.000485\). In absolute terms, the
first distortion is 1.76 times the second.

\textbf{Table 10.} \emph{Aggregation distortion under two
constructions.}

\begin{longtable}[]{@{}
  >{\raggedright\arraybackslash}p{(\columnwidth - 4\tabcolsep) * \real{0.3333}}
  >{\raggedright\arraybackslash}p{(\columnwidth - 4\tabcolsep) * \real{0.3333}}
  >{\raggedright\arraybackslash}p{(\columnwidth - 4\tabcolsep) * \real{0.3333}}@{}}
\toprule\noalign{}
\begin{minipage}[b]{\linewidth}\raggedright
Quantity
\end{minipage} & \begin{minipage}[b]{\linewidth}\raggedright
Conventional construction
\end{minipage} & \begin{minipage}[b]{\linewidth}\raggedright
Identity-consistent
\end{minipage} \\
\midrule\noalign{}
\endhead
\bottomrule\noalign{}
\endlastfoot
Pooled indirect effect & 0.008744 & 0.008744 \\
Decomposed target & 0.009598 & 0.009229 \\
Aggregation distortion (AD) & -0.000854 & -0.000485 \\
Percent distortion (PD) & 8.90\% & 5.26\% \\
Absolute AD ratio & 1.76 & 1.00 \\
\end{longtable}

The comparison is diagnostic. The conventional construction changes both
the order of covariate adjustment and the weighting of the between-level
regression, so Table 10 does not isolate one source. It shows something
more basic: the same pooled estimate can lead to a different reported
decomposition when the weights and paths are not defined on the same
scale. The identity check in Table 9 catches that mismatch before
\(AD\) or \(PD\) is interpreted.

The simulation in Section 14.5 shows the same pattern across a range of
designs. The conventional construction gives larger absolute distortion
in every cell of that grid, with ratios from 1.05 to 2.45. The ratio is
not monotone in cluster-size imbalance, so the result should not be read
as a simple dose--response pattern. The practical lesson is narrower:
weights and level-specific paths need to be constructed so that they
reproduce the pooled identity.

The example also shows why these diagnostics should be kept separate.
The two design weights differ clearly. The between-school \(b\)-path is
larger at the point estimate, but its difference from the within-school
path is uncertain. The point estimate of \(PD\) is 5.26\%, yet the
signed \(AD\) interval includes zero. Person- and cluster-weighted
between indirect effects are close. These statements answer different
questions; none of them alone is a general verdict about whether
clustering matters.

\subsection{15.2 Uncertainty and Scope}\label{uncertainty-and-scope}

The within-school indirect effect is 0.00901 (95\% CI {[}0.00575,
0.01233{]}), and the person-weighted between-school indirect effect is
0.00980 (95\% CI {[}0.00283, 0.01970{]}). The pooled effect is 0.00874
(95\% CI {[}0.00584, 0.01174{]}), and the decomposed target is 0.00923
(95\% CI {[}0.00609, 0.01265{]}). Their signed difference is
\(AD=-0.000485\) with a 95\% CI of {[}-0.001959, 0.000343{]}. Thus the
data do not establish nonzero population distortion. The weight
difference is much clearer: \(\Delta=-0.173\) (95\% CI {[}-0.193,
-0.151{]}). At \(r=0.50\), the standardized effect is 0.00940 (95\% CI
{[}0.00531, 0.01443{]}), and \(DSI_2=0.278\) (95\% CI {[}0.094,
0.750{]}). \(PD\) is reported only as a descriptive point estimate.
Because \(PD=100|AD|/|IE_{\text{dec}}|\), taking absolute values folds
the bootstrap distribution at zero; when the signed \(AD\) interval
crosses zero, a positive lower percentile for \(|AD|\) would be
misleading.

\begin{figure}[htbp]
\centering
\includegraphics[width=0.72\textwidth]{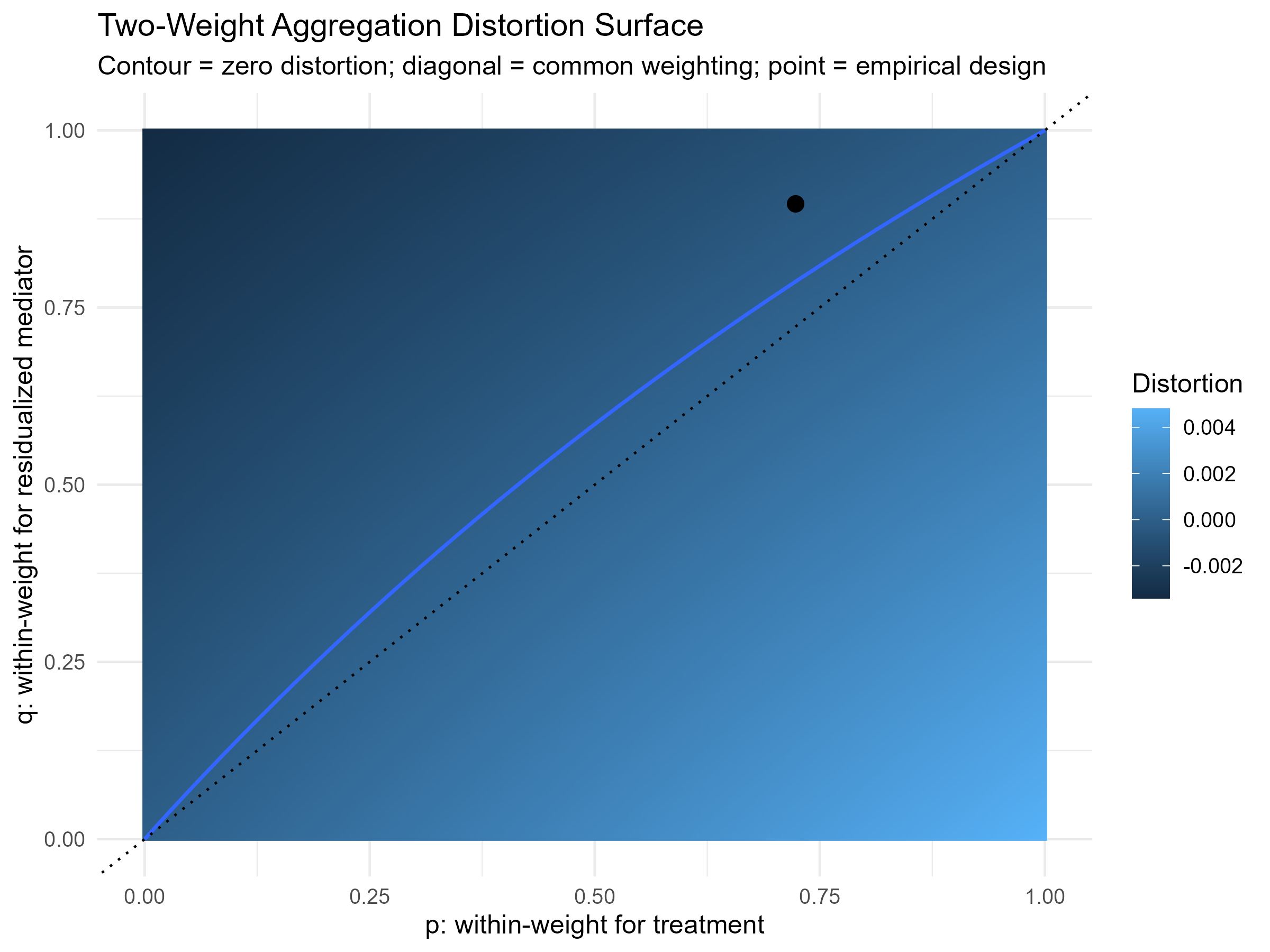}
\caption{Aggregation distortion over the design square for the ECLS-K:2011 baseline-school application. The realized design is \((p,q)=(0.723,0.896)\), so the two within-level weights differ by 0.173. At that design the pooled indirect effect lies below the decomposed target (\(AD=-0.000485\)). The zero-distortion contour lies away from the empirical point; over the stated \(\pm0.10\) design rectangle the pooled effect remains positive while the distortion bound spans zero.}
\label{fig:ecls-surface}
\end{figure}

Figure~\ref{fig:ecls-surface} places the observed ECLS-K design on the
two-weight distortion surface and shows its position relative to the
equal-weight diagonal.

\begin{figure}[htbp]
\centering
\includegraphics[width=0.72\textwidth]{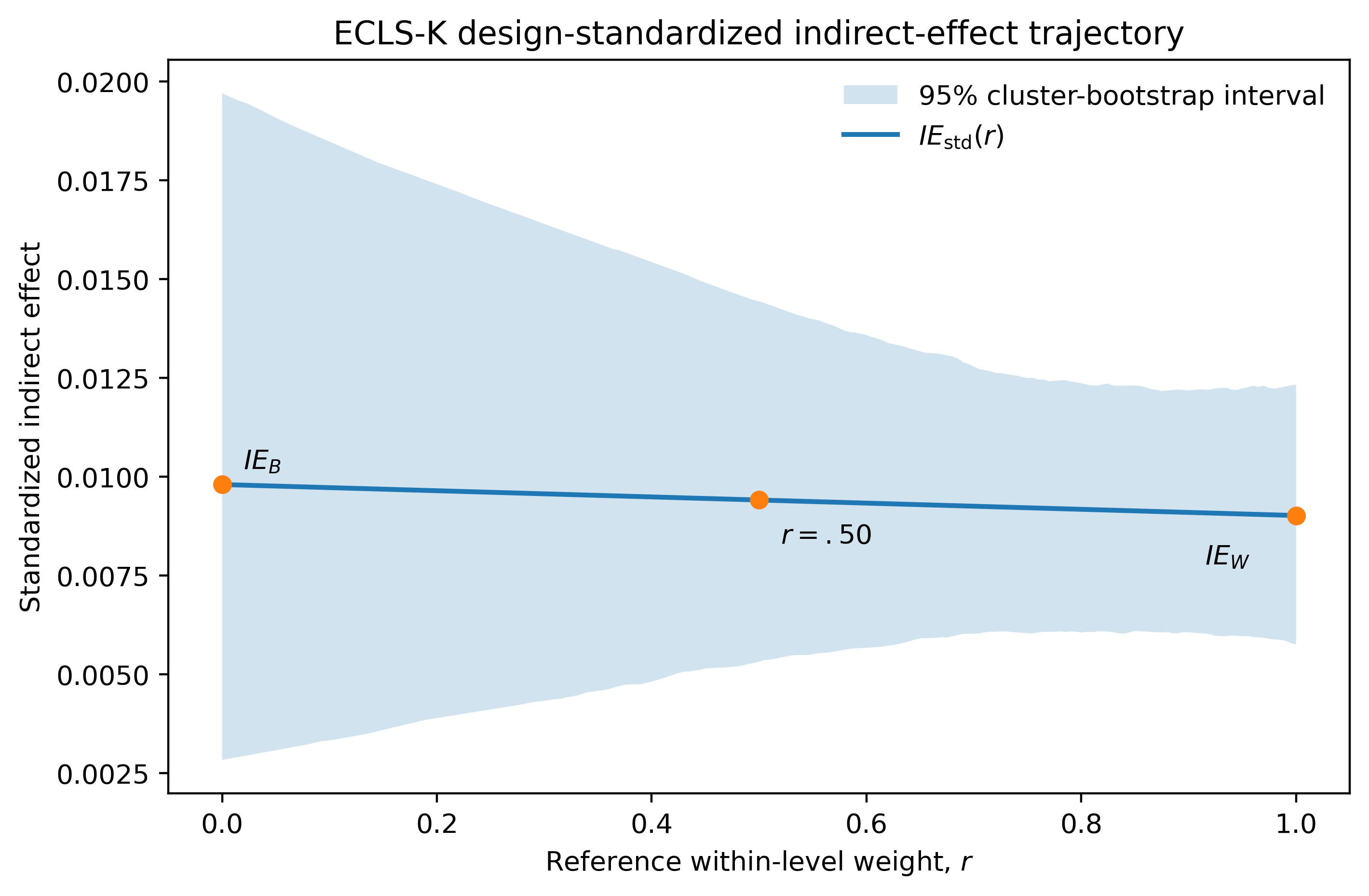}
\caption{Design-standardized indirect-effect trajectory for the ECLS-K:2011 illustration. The line is \(IE_{std}(r)=rIE_W+(1-r)IE_B\); the band is the pointwise 95\% school-cluster bootstrap interval computed from the same 1,000 bootstrap replications used in Table 7. The endpoints are the between- and within-baseline-school indirect effects, and the midpoint marks the prespecified \(r=0.50\) summary.}
\label{fig:ecls-standardized}
\end{figure}

Figure~\ref{fig:ecls-standardized} shows the full standardized
trajectory over \(r\in[0,1]\). The point estimate changes little across
the range, while the interval widens toward the between-school endpoint
because the between-level effect is estimated less precisely.

The two level-specific indirect effects have the same sign at the point
estimate, so there is no cross-level suppression. Only 1 of the 1,000
bootstrap samples showed suppression, and only 1 produced an admissible
sign-reversal frontier. For that reason, \(SPI\) is not reported.

The example has several limits. Assumption 1 is not established by the
observational design. The analysis is complete-case and does not use the
NCES complex survey weights. The public-use file suppresses the original
sampled PSU identifier (PSUID) for confidentiality (Tourangeau et
al., 2019), so a PSU-level bootstrap cannot be
implemented from that file. We instead resample the
spring-kindergarten baseline schools that define the decomposition.
Because \texttt{S2\_ID} is a baseline-school identifier, the contextual
interpretation is tied to those baseline cohorts. These limitations
restrict substantive interpretation of the coefficients but do not
change the algebraic identities illustrated here.

\section{16. Practical Implementation}\label{practical-implementation}

The algebra in Sections 5--11 assumes that the weights and
level-specific paths refer to the same residualized projection. Sections
14.5 and 15.1 show what happens when they do not. The following steps
give the construction used throughout the paper.

\subsection{16.1 The Construction}\label{the-construction}

\textbf{Step 1. Residualize at the individual level.} Regress the
treatment, mediator, and outcome on the covariates using the pooled
individual-level data and keep the residuals. This step comes before the
within--between decomposition. Decomposing the covariates first, or
adjusting for them separately within levels, changes the projection and
can break Lemma 2.

\textbf{Step 2. Decompose the residualized treatment.} Form the cluster
means and within-cluster deviations of the treatment residuals. Compute
the variance weights from those same residuals. Raw-treatment weights
do not describe the covariate-adjusted pooled regression. In Section 15,
mixing raw-scale weights with separately adjusted paths changes the
reported \(PD\) from 5.26\% to 8.90\%.

\textbf{Step 3. Estimate the \(a\)-paths with person weighting.}
Project the residualized mediator on the within- and between-components
of residualized treatment. The between-level projection must weight each
cluster by \(n_j\) if it is to match the coefficient mixed by the pooled
individual-level regression. An unweighted cluster-means regression
targets a different estimand.

\textbf{Step 4. Repeat the construction for the \(b\)-path.}
Residualize the mediator and outcome on treatment, decompose the
residualized mediator, and estimate \(\widehat q\),
\({\widehat{\beta}}_{2W}\), and \({\widehat{\beta}}_{2B}\).

\textbf{Step 5. Check the pooled identities.} Verify

\[\widehat{p}\,{\widehat{\alpha}}_{W} + \left( 1 - \widehat{p} \right)\,{\widehat{\alpha}}_{B} = {\widehat{\alpha}}_{\text{pool}},\quad\quad\widehat{q}\,{\widehat{\beta}}_{2W} + \left( 1 - \widehat{q} \right)\,{\widehat{\beta}}_{2B} = {\widehat{\beta}}_{2,\text{pool}}.\]

These are exact algebraic identities for the construction in Definition
12. The residual should therefore be at machine precision, not merely
small after rounding. A visible error means that the weights and paths
are not on the same projection scale, so downstream distortion measures
should not be interpreted. In the most extreme cell of Table 6, the
conventional path identity error exceeds the fitted pooled coefficient
itself.

\subsection{16.2 What to Report}\label{what-to-report}

A concise applied report should include the following items.

\begin{enumerate}
\def\labelenumi{\arabic{enumi}.}
\tightlist
\item
  Sample structure: \(J\), \(N\), the harmonic mean cluster size, the
  maximum, and \(CV\left( n_{j} \right)\). The last is the single best
  predictor of whether the scope condition of Proposition 4 will bind.
\item
  The identity check of Step 5, reported as a residual rather than
  asserted. A single line stating that both identities held to
  \(10^{- 16}\) is sufficient and should become conventional.
\item
  Design coordinates \(\left( \widehat{p},\widehat{q} \right)\), the
  discrepancy \(\widehat{\Delta}\), and the distance from the
  equal-weight diagonal.
\item
  If Proposition 1 is illustrated empirically, report the
  between-cluster weight and the ICC on the same treatment scale and
  state that scale explicitly. Do not compare a raw-treatment ICC with a
  covariate-residualized design weight as though they were commensurate.
  The design coordinates used in the pooled identity must be computed
  after the covariate residualization required by Remark 5.
\item
  Level-specific paths and indirect effects with cluster-bootstrap
  intervals, both person- and cluster-weighted at the between level, and
  their difference as an informative-cluster-size diagnostic.
\item
  \({\widehat{IE}}_{\text{pool}}\), \({\widehat{IE}}_{\text{dec}}\),
  \(\widehat{AD}\), \(\widehat{PD}\), and the split into
  \({\widehat{A}}_{nl}\) and \({\widehat{A}}_{wt}\).
\item
  \(IE_{\text{std}}(r)\) for a prespecified \(r\), with the trajectory
  and slope of Section 10.4.
\item
  Geometry: the Hessian eigenvalues
  \(\pm \left| \Delta_{\alpha}\Delta_{\beta} \right|\) as a check that
  the surface is a saddle, \(DSI_{2}\), and the sharp bounds
  \(\mathcal{R}_{P}\) and \(\mathcal{R}_{A}\) over a stated design
  region. Report \(SPI\) only when a frontier exists, with the bootstrap
  frontier-existence frequency.
\end{enumerate}

\subsection{16.3 Common Failure Modes}\label{common-failure-modes}

The same mistakes tend to recur, and each can be checked directly.

\emph{Using the ICC as the between-cluster weight.} Under balance, the
difference is \((1-ICC)/n\) (Proposition 1). With \(ICC=0.05\) and
\(n=10\), the correct weight is almost three times the ICC. The check is
simply to report both quantities.

\emph{Computing weights from the raw treatment after covariate
adjustment.} The Step 5 identity check will reveal the mismatch.

\emph{Using unweighted cluster-means slopes in the pooled
decomposition.} Such slopes can be useful contextual estimands, but they
are not generally the between-level coefficients mixed by the pooled
individual-level regression. The identity check detects the difference,
even when the two indirect-effect products happen to be close.

\emph{Using the wrong cluster identifier.} In longitudinal data, the
cluster definition is part of the estimand. A late-wave identifier can
refer to a context measured after the mediator or outcome, and status
codes can create artificial ``clusters.'' Before fitting the model,
check the codebook, the timing of the identifier, the largest cluster
sizes, and the sample-flow table. If several pre-outcome cluster
definitions are substantively reasonable, they can be compared as an
estimand-sensitivity analysis. A post-outcome identifier should not be
treated as an equally valid alternative.

\emph{Treating a small \(\widehat{AD}\) as proof that pooling is
safe.} The two components can cancel, and a design with \(A\approx0\)
may simply lie near the zero-distortion frontier. Reporting
\(\mathcal R_A\) and \(DSI_2\) shows whether the conclusion is stable to
changes in the design weights.

\section{17. Discussion}\label{discussion}

\subsection{17.1 Main Findings}\label{what-the-framework-establishes}

A pooled product-of-coefficients analysis in clustered data estimates a
specific design-dependent quantity. The pooled \(a\)-path and \(b\)-path
are generally mixed with different within--between variance weights.
Theorem 1 gives the resulting probability limit. Theorem 2 then shows
how its difference from a level-respecting target separates into two
parts: a nonlinearity component and a weight-mismatch component.

The weight mismatch changes some familiar intuition. Under equal
weights, the formula looks symmetric in the two mediation paths. With
unequal weights, equal \(b\)-paths eliminate distortion, but equal
\(a\)-paths do not. The reason is that the \(b\)-path is mixed
differently in the pooled model and in the decomposed reference target.

The two-weight geometry makes this dependence easy to see. Distortion is
bilinear in the treatment and residualized-mediator weights. Its
curvature comes entirely from their interaction, its largest absolute
values occur at the corners of the design square, and its zero set is a
frontier rather than a whole region. These properties also make design
uncertainty easy to study: four corner evaluations give sharp bounds
over any rectangular range of plausible weights.

The ECLS-K illustration shows why the construction details matter in
practice. Person- and cluster-weighted between-school indirect effects
are close, but the cluster-weighted paths do not reproduce the fitted
pooled coefficients. The combined conventional construction gives an
absolute distortion 1.76 times the identity-consistent value. The signed
identity-consistent \(AD\) interval includes zero, so this difference is
best read as sensitivity of the point estimate to the construction, not
as evidence of nonzero population distortion. A simple pooled-identity
check catches the problem before \(AD\) or \(PD\) is interpreted.

The design-standardized effect offers a separate summary for comparison
across studies. It fixes the reference weight in advance instead of
using the realized design. The within- and between-level effects still
need to be reported, especially when they differ strongly, but the
standardized mixture no longer changes simply because the cluster-size
distribution changes.

\subsection{17.2 Limitations}\label{limitations}

The framework does not solve causal identification. In observational
data, Assumption 1 remains strong and untestable. If it is not credible,
the results describe relationships among projection parameters rather
than causal effects.

The observed-projection convention is also important. The between-level
paths are defined using observed cluster means. A latent cluster-level
mechanism is a different target and can be affected by measurement error
when cluster sizes are bounded. Proposition 3 gives a simple reliability
result for the \(a\)-path, but the \(b\)-path contains two error-prone
between-level regressors and generally needs a multivariate
errors-in-variables treatment.

The model is linear, has constant path coefficients, and contains no
treatment--mediator interaction. Under nonlinear links or interactions,
the product-of-coefficients representation changes and the bilinear
geometry need not survive. \(PD\) is unstable when its denominator is
close to zero, and \(SPI\) is nonregular because the existence of a
sign-reversal frontier is itself estimated. Between-level interval
coverage is also somewhat low at small numbers of clusters; a
small-sample correction for that setting remains to be studied.

Finally, the contextual between-level estimand assumes that a shift in a
cluster mean is a meaningful contrast for the application. That
interpretation will not be appropriate in every setting.

\subsection{17.3 Extensions}\label{extensions}

Several extensions are immediate. Random \(a\)- and \(b\)-paths would
allow direct study of the distribution of cluster-specific indirect
effects, not only their mean. Nonlinear links and
treatment--mediator interactions would replace the simple product with
marginal indirect-effect functionals; whether a useful analogue of the
two-weight geometry exists in those models is an open question.

Informative cluster size also deserves more work. Person- and
cluster-weighted between-level effects answer different questions, and a
formal framework for choosing between them would be useful. A related
direction is sensitivity analysis for partial knowledge of the
cross-level path covariance in Corollary 1.

\section{Appendix A: Regularity
Conditions}\label{appendix-a-regularity-conditions}

\textbf{(A1) Cluster independence.} The cluster-level data are i.i.d.
across \(j = 1,\ldots,J\).

\textbf{(A2) Within-cluster independence.} Conditional on \(j\), the
residuals \(\{\left( u_{ij},e_{ij} \right)\}\) are i.i.d. across \(i\)
with conditional mean zero.

\textbf{(A3) Bounded moments.} \(\mathbb{E}\left| D_{ij} \right|^{4}\),
\(\mathbb{E}\left| M_{ij} \right|^{4}\),
\(\mathbb{E}\left| Y_{ij} \right|^{4} < \infty\).

\textbf{(A4) Non-degeneracy.} \(Var\left( D_{ij}^{W} \right) > 0\) and
\(Var\left( {\bar{D}}_{j} \right) > 0\), ensuring strictly positive
variance weights.

\textbf{(A5) Cluster-size regularity.}
\(n_{\min} \leq n_{j} \leq n_{\max} < \infty\) for the bounded-size
asymptotic sequence.

\textbf{(A6) Linear projection conditions.} The population linear
projections of \(M\) on \(\left( D,\mathbf{X} \right)\) and of \(Y\) on
\(\left( D,M,\mathbf{X} \right)\) exist with nonsingular second-moment
matrices, and the residualized mediator admits the orthogonal
within--between decomposition of Section 3.1.

\section{Appendix B: Proofs}\label{appendix-b-proofs}

\textbf{Proof of Lemma 1.} (a) Within each cluster
\(\sum_{i}^{}\left( D_{ij} - {\bar{D}}_{j} \right) = 0\) by construction
of the sample mean, for any \(n_{j}\). Hence
\(\sum_{j}^{}{\sum_{i}^{}{\bar{D}}_{j}}D_{ij}^{W} = \sum_{j}^{}{\bar{D}}_{j} \cdot 0 = 0\).
In population,
\(\mathbb{E}\left\lbrack {\bar{D}}_{j}D_{ij}^{W} \right\rbrack = \mathbb{E}\left\lbrack {\bar{D}}_{j}\,\mathbb{E}\left( D_{ij} - {\bar{D}}_{j} \mid j \right) \right\rbrack = 0\),
and since \(\mathbb{E}\left\lbrack D_{ij}^{W} \right\rbrack = 0\) the
covariance vanishes. Additivity of variances and of the weights follows.
(b) For \(n_{j} = n\), \({\bar{D}}_{j} = \mu + B_{j} + {\bar{W}}_{j}\)
gives
\(Var\left( {\bar{D}}_{j} \right) = \sigma_{B}^{2} + \sigma_{W}^{2}/n\),
and \(Var\left( D_{ij}^{W} \right) = \sigma_{W}^{2}(1 - 1/n)\); dividing
by \(\sigma_{B}^{2} + \sigma_{W}^{2}\) and writing \(\rho = ICC(D)\)
yields \(\omega_{B}(D) = \rho + (1 - \rho)/n\). (c) Under imbalance
\(Var\left( {\bar{D}}_{j} \right)\) averages \(\sigma_{W}^{2}/n_{j}\)
across clusters, which the harmonic mean approximates. \(\blacksquare\)

\textbf{Proof of Proposition 1.} From Lemma 1(b),
\(\omega_{B}(D) = \rho + (1 - \rho)/n = \left\lbrack 1 + (n - 1)\rho \right\rbrack/n\),
and \(\omega_{B}(D) - \rho = (1 - \rho)/n > 0\), decreasing in \(n\) and
in \(\rho\). \(\blacksquare\)

\textbf{Proof of Proposition 2.} Under Definition 1 with no
treatment--mediator interaction, substituting the structural equations
into Definition 2 gives \(M(b,w + 1) - M(b,w) = \alpha_{W}\), a purely
within-cluster difference entering \(Y\) through \(\beta_{2W}\), so
\(NIE_{W} = \alpha_{W}\beta_{2W}\). Under Assumption 1(W1)--(W2) and
Assumption 2 the corresponding conditional expectations are identified
from observed data, and linearity makes the cross-world quantity equal
the observed-data functional. Similarly
\(M(b + 1,w) - M(b,w) = \alpha_{B}\) enters through \({\bar{M}}_{j}\)
and hence \(\beta_{2B}\), giving \(NIE_{B} = \alpha_{B}\beta_{2B}\).
Direct effects follow by holding the mediator argument fixed.
\(\blacksquare\)

\textbf{Proof of Proposition 3.} With the mediator equation on latent
components,
\({\bar{M}}_{j} = \alpha_{B}^{\dagger}B_{j} + \zeta_{j}^{M} + {\bar{u}}_{j}\)
and \({\bar{D}}_{j} = \mu + B_{j} + {\bar{W}}_{j}\). Since
\(B_{j}\bot{\bar{W}}_{j}\),
\(Cov\left( {\bar{M}}_{j},{\bar{D}}_{j} \right) = \alpha_{B}^{\dagger}\sigma_{B}^{2}\)
while
\(Var\left( {\bar{D}}_{j} \right) = \sigma_{B}^{2} + \sigma_{W}^{2}/n\),
so the projection coefficient is \(\lambda_{n}\alpha_{B}^{\dagger}\).
Dividing numerator and denominator by
\(\sigma_{B}^{2} + \sigma_{W}^{2}\) gives the ICC form. As
\(n \rightarrow \infty\), \(\lambda_{n} \rightarrow 1\); for fixed \(n\)
the limit does not depend on \(J\). \(\blacksquare\)

\textbf{Proof of Proposition 4.} Without covariates,
\({\widehat{\alpha}}_{\text{pool}} = \widehat{Cov}(D,M)/\widehat{Var}(D)\)
with all moments taken over the \(N\) individual observations. By Lemma
1(a),
\(\widehat{Cov}(D,M) = \widehat{Cov}\left( {\bar{D}}_{j},{\bar{M}}_{j} \right) + \widehat{Cov}\left( D^{W},M^{W} \right)\),
where the first moment is again taken over individuals and therefore
weights each cluster by \(n_{j}\). Dividing,
\({\widehat{\alpha}}_{\text{pool}} = {\widehat{\omega}}_{W}{\widehat{\alpha}}_{W} + {\widehat{\omega}}_{B}{\widehat{\alpha}}_{B}^{per}\)
exactly, where \({\widehat{\alpha}}_{B}^{per}\) is the
\(n_{j}\)-weighted cluster-means slope. Part (a) is immediate since with
\(n_{j} = n\) the weights are constant. Part (b) follows from the
standard weighted-versus-unweighted least-squares difference formula
applied to the \(J\) cluster-level rows. \(\blacksquare\)

\textbf{Proof of Lemma 2.} By Lemma 1(a),
\(Cov(D,M) = \alpha_{W}Var\left( D^{W} \right) + \alpha_{B}Var\left( \bar{D} \right)\),
so
\(\alpha_{\text{pool}} = \omega_{W}(D)\alpha_{W} + \omega_{B}(D)\alpha_{B}\).
For the \(b\)-path, apply Frisch--Waugh--Lovell: with
\(M^{\bot} = M - proj\left( M \mid D,\mathbf{X} \right)\), the
orthogonal within--between decomposition of \(M^{\bot}\) gives
\(\beta_{2,\text{pool}} = \omega_{W}(M \mid D)\beta_{2W} + \omega_{B}(M \mid D)\beta_{2B}\).
\(\blacksquare\)

\textbf{Proof of Theorem 1.} Lemma 2 gives both pooled path limits; the
continuous mapping theorem gives
\(plim\left( {\widehat{IE}}_{\text{pool}} \right) = plim\left( {\widehat{\alpha}}_{\text{pool}} \right)plim\left( {\widehat{\beta}}_{2,\text{pool}} \right)\).
\(\blacksquare\)

\textbf{Proof of Theorem 2.} Write
\({\bar{\beta}}_{D} = \omega_{W}(D)\beta_{2W} + \omega_{B}(D)\beta_{2B}\)
and
\({\bar{\beta}}_{M} = \omega_{W}(M \mid D)\beta_{2W} + \omega_{B}(M \mid D)\beta_{2B}\),
so Theorem 1 reads
\(plim\left( IE_{\text{pool}} \right) = \bar{\alpha}{\bar{\beta}}_{M}\).
Add and subtract \(\bar{\alpha}{\bar{\beta}}_{D}\):

\[A = \left( \bar{\alpha}{\bar{\beta}}_{D} - IE_{\text{dec}} \right) + \bar{\alpha}\left( {\bar{\beta}}_{M} - {\bar{\beta}}_{D} \right).\]

For the first bracket, put \(p = \omega_{W}(D)\) and expand
\(\bar{\alpha}{\bar{\beta}}_{D} = p^{2}\alpha_{W}\beta_{2W} + p(1 - p)\left( \alpha_{W}\beta_{2B} + \alpha_{B}\beta_{2W} \right) + (1 - p)^{2}\alpha_{B}\beta_{2B}\).
Using \(p^{2} = p - p(1 - p)\) and \((1 - p)^{2} = (1 - p) - p(1 - p)\),

\[\bar{\alpha}{\bar{\beta}}_{D} = p\, IE_{W} + (1 - p)\, IE_{B} + p(1 - p)\left\lbrack \alpha_{W}\beta_{2B} + \alpha_{B}\beta_{2W} - \alpha_{W}\beta_{2W} - \alpha_{B}\beta_{2B} \right\rbrack,\]

and the bracket factors as \(- \Delta_{\alpha}\Delta_{\beta}\). Since
\(IE_{\text{dec}} = p\, IE_{W} + (1 - p)IE_{B}\), the first bracket
equals \(A_{nl}\). For the second, since \(\omega_{W} = 1 - \omega_{B}\)
in both systems,
\({\bar{\beta}}_{M} - {\bar{\beta}}_{D} = - \Delta\beta_{2W} + \Delta\beta_{2B} = - \Delta\Delta_{\beta}\),
so
\(\bar{\alpha}\left( {\bar{\beta}}_{M} - {\bar{\beta}}_{D} \right) = A_{wt}\).

\emph{Numerical verification.} With \(\alpha_{W} = 0.30\),
\(\alpha_{B} = 0.50\), \(\beta_{2W} = 0.40\), \(\beta_{2B} = 0.70\),
\(\omega_{B}(D) = 0.2929\), \(\omega_{B}(M \mid D) = 0.3251\):
\(\bar{\alpha} = 0.35858\), \(\Delta = + 0.0322\),
\(A_{nl} = - 0.012427\), \(A_{wt} = + 0.003464\), \(A = - 0.008963\).
Direct computation gives
\(plim\left( IE_{\text{pool}} \right) = 0.178404\) and
\(IE_{\text{dec}} = 0.187367\), differing by \(- 0.008963\).
\(\blacksquare\)

\textbf{Proof of Corollary 1.} For the two-point distribution with
masses \(\omega_{W}(D),\omega_{B}(D)\),
\(\mathbb{E}_{\omega}\left( \alpha_{L} \right)\mathbb{E}_{\omega}\left( \beta_{L} \right) - \mathbb{E}_{\omega}\left( \alpha_{L}\beta_{L} \right) = - {Cov}_{\omega}\left( \alpha_{L},\beta_{L} \right) = - \omega_{W}\omega_{B}\Delta_{\alpha}\Delta_{\beta} = A_{nl}\).
When \(\Delta = 0\), \(A_{wt} = 0\). \(\blacksquare\)

\textbf{Proof of Proposition 5.} (a) and (b) follow from the
factorization of \(A_{nl}\) and
\(\max_{\omega \in \lbrack 0,1\rbrack}\omega(1 - \omega) = 1/4\). (c)
follows by dividing \(\left| A_{wt} \right|\) by
\(\left| A_{nl} \right|\); the common factor
\(\left| \Delta_{\beta} \right|\) cancels and the quotient diverges as
\(\left| \Delta_{\alpha} \right| \rightarrow 0\). Boundedness of each
term separately follows from \(|\Delta| \leq 1\) and
\(\omega_{W}\omega_{B} \leq 1/4\). (d) sets \(A_{nl} + A_{wt} = 0\) and
solves for \(\Delta\). \(\blacksquare\)

\textbf{Proof of Corollary 3.} With a \(K\)-point weight distribution,
\(plim\left( IE_{\text{pool}} \right) = \left( \sum_{\ell}^{}\omega_{\ell}\alpha_{\ell} \right)\left( \sum_{\ell}^{}\omega_{\ell}^{M}\beta_{\ell} \right)\)
and
\(IE_{\text{dec}} = \sum_{\ell}^{}\omega_{\ell}\alpha_{\ell}\beta_{\ell}\).
Adding and subtracting
\(\bar{\alpha}\sum_{\ell}^{}\omega_{\ell}\beta_{\ell}\) separates the
difference into
\(- {Cov}_{\omega}\left( \alpha_{\ell},\beta_{\ell} \right)\) and
\(- \bar{\alpha}\sum_{\ell}^{}\left( \omega_{\ell}^{M} - \omega_{\ell} \right)\beta_{\ell}\).
\(\blacksquare\)

\textbf{Proof of Theorem 3.} \(a(p)\) is affine in \(p\) and \(b(q)\) is
affine in \(q\), so \(P = a(p)b(q)\) is affine in each argument
separately, hence bilinear. Differentiating,
\(\partial P/\partial p = a^{\prime}(p)b(q) = \Delta_{\alpha}b(q)\) and
\(\partial P/\partial q = a(p)b^{\prime}(q) = \Delta_{\beta}a(p)\); since
\(a^{\prime}\) and \(b^{\prime}\) are constants,
\(\partial^{2}P/\partial p^{2} = \partial^{2}P/\partial q^{2} = 0\) and
\(\partial^{2}P/\partial p\partial q = a^{\prime}(p)b^{\prime}(q) = \Delta_{\alpha}\Delta_{\beta}\).
\(T(p) = p\, IE_{W} + (1 - p)IE_{B}\) is affine in \(p\) with
\(T^{\prime}(p) = IE_{W} - IE_{B}\) and does not involve \(q\), so
\(A = P - T\) has the stated derivatives and the same cross-partial.
\(\blacksquare\)

\textbf{Proof of Corollary 4.} Setting \(A(p,q) = 0\) gives
\(a(p)b(q) = T(p)\). If \(a(p) \neq 0\), then \(b(q) = T(p)/a(p)\), and
since \(b(q) = \beta_{2B} + q\Delta_{\beta}\) with
\(\Delta_{\beta} \neq 0\), solving for \(q\) gives \(q_{0}(p)\). At
\(p = 0\): \(T(0) = IE_{B} = \alpha_{B}\beta_{2B}\) and
\(a(0) = \alpha_{B}\), so \(b\left( q_{0} \right) = \beta_{2B}\) and
\(q_{0}(0) = 0\). At \(p = 1\): \(T(1) = \alpha_{W}\beta_{2W}\),
\(a(1) = \alpha_{W}\), so \(b\left( q_{0} \right) = \beta_{2W}\) and
\(q_{0}(1) = 1\). Continuity of \(A\) and the sign change across
\(\mathcal{Z}\) give the partition. If \(\Delta_{\beta} = 0\) then
\(b(q) \equiv \beta_{2B}\) and
\(A(p,q) = a(p)\beta_{2B} - T(p) = - p(1 - p)\Delta_{\alpha} \cdot 0 = 0\)
identically. \(\blacksquare\)

\textbf{Proof of Theorem 4.} A bilinear function on a rectangle attains
its extrema at the vertices, since for fixed \(q\) it is affine in \(p\)
and therefore maximized at an endpoint, and likewise in \(q\). The
corner values follow by direct substitution:
\(A(0,0) = \alpha_{B}\beta_{2B} - IE_{B} = 0\);
\(A(1,1) = \alpha_{W}\beta_{2W} - IE_{W} = 0\);
\(A(0,1) = \alpha_{B}\beta_{2W} - \alpha_{B}\beta_{2B} = \alpha_{B}\Delta_{\beta}\);
\(A(1,0) = \alpha_{W}\beta_{2B} - \alpha_{W}\beta_{2W} = - \alpha_{W}\Delta_{\beta}\).
The supremum of \(|A|\) is therefore
\(\max\left( \left| \alpha_{B}\Delta_{\beta} \right|,\left| \alpha_{W}\Delta_{\beta} \right| \right) = \left| \Delta_{\beta} \right|\max\left( \left| \alpha_{W} \right|,\left| \alpha_{B} \right| \right)\).
\(\blacksquare\)

\textbf{Proof of Corollary 5.} \(P\) restricted to
\(\left\lbrack p_{L},p_{U} \right\rbrack \times \left\lbrack q_{L},q_{U} \right\rbrack\)
is bilinear on a rectangle, so by the argument of Theorem 4 its extrema
are attained at the four vertices; the bounds are therefore attained and
sharp. The same applies to \(A\), which is bilinear by Theorem 3. The
sign statements follow because \(\mathcal{R}\) is an interval containing
all attainable values. \(\blacksquare\)

\textbf{Proof of Theorem 5.} Setting \(p = q = \omega\) in \(P\) gives
\(g(\omega) = \left( \alpha_{B} + \omega\Delta_{\alpha} \right)\left( \beta_{2B} + \omega\Delta_{\beta} \right) = \alpha_{B}\beta_{2B} + \omega\left( \alpha_{B}\Delta_{\beta} + \beta_{2B}\Delta_{\alpha} \right) + \omega^{2}\Delta_{\alpha}\Delta_{\beta}\),
a quadratic with leading coefficient \(\Delta_{\alpha}\Delta_{\beta}\)
and \(g(0) = IE_{B}\), \(g(1) = IE_{W}\). Setting
\(g^{\prime}(\omega) = 0\) gives \(\omega_{v}\). A quadratic is monotone on
an interval if and only if its stationary point lies outside it; if
\(\omega_{v} \in (0,1)\) then \(g\left( \omega_{v} \right)\) is a strict
extremum on \(\lbrack 0,1\rbrack\), and since the endpoint values are
\(IE_{B}\) and \(IE_{W}\) it lies strictly outside their range according
to the sign of \(\Delta_{\alpha}\Delta_{\beta}\). Part (d) is immediate.
\(\blacksquare\)

\textbf{Proof of Proposition 6.} A product vanishes if and only if a
factor does. \(a(p) = 0\) gives
\(p^{*} = \alpha_{B}/\left( \alpha_{B} - \alpha_{W} \right)\), which
lies in \((0,1)\) exactly when \(\alpha_{W}\) and \(\alpha_{B}\) have
opposite signs; identically for \(b(q)\). \(\blacksquare\)

\textbf{Proof of Proposition 7.} \({\widehat{IE}}_{\text{std}}(r)\) is a
fixed linear combination of two asymptotically normal estimators, so its
variance is \(r^{2}V_{W} + (1 - r)^{2}V_{B} + 2r(1 - r)C_{WB}\). Under
balance and joint normality, cluster means and within-cluster deviations
are independent, so the between-level and within-level estimating
equations depend on independent data and \(C_{WB} = 0\).
\(\blacksquare\)

\textbf{Proof of Proposition 8.} Under (i),
\(Cov(D,M) = \mathbb{E}\left\lbrack \left( \alpha_{W} + a_{j} \right)D_{ij}^{W}D_{ij} \right\rbrack + \alpha_{B}Var\left( {\bar{D}}_{j} \right)\)
and
\(\mathbb{E}\left\lbrack a_{j}D_{ij}^{W}D_{ij} \right\rbrack = \mathbb{E}\left\lbrack a_{j} \right\rbrack\mathbb{E}\left\lbrack D_{ij}^{W}D_{ij} \right\rbrack = 0\),
so the pooled \(a\)-path limit is exactly as in Lemma 2. For the
\(b\)-path the numerator involves
\(\mathbb{E}\left\lbrack \beta_{2j}\left( M_{ij}^{W} \right)^{2} \right\rbrack\);
because \(M_{ij}^{W}\) depends on the random paths only through
\(a_{j}\), this equals
\(\beta_{2W}\mathbb{E}\left\lbrack \left( M_{ij}^{W} \right)^{2} \right\rbrack + Cov\left( b_{j},a_{j}^{2} \right)Var\left( D_{ij}^{W} \right)\),
and the second term vanishes by (ii). The within-level estimand is
\(\mathbb{E}\left\lbrack \alpha_{j}\beta_{2j} \right\rbrack = IE_{W} + \tau_{ab}\),
so \(IE_{\text{dec}}^{*} = IE_{\text{dec}} + \omega_{W}(D)\tau_{ab}\)
and \(A^{*} = A - \omega_{W}(D)\tau_{ab}\). Substituting Theorem 2 for
\(A\) gives the three-term expression. \(\blacksquare\)

\textbf{Proof of Theorem 7.} Under (A1)--(A6) each level-specific path
estimator is a consistent M-estimator for its linear-projection target
on observed cluster means, based on independent clusters; the continuous
mapping theorem gives consistency of both products. A cluster-level
central limit theorem applied to the stacked estimating equations gives
\(\sqrt{J}\left( {\widehat{\mathbf{\theta}}}_{W} - \mathbf{\theta}_{W} \right)\overset{d}{\rightarrow}\mathcal{N}\left( \mathbf{0},\mathbf{\Sigma}_{W} \right)\).
Since \(g\left( \mathbf{\theta} \right) = \theta_{1}\theta_{2}\) is
continuously differentiable with
\(\nabla g = \left( \theta_{2},\theta_{1} \right)^{\prime}\), the delta
method yields \(V_{W} = \nabla g^{\prime}\mathbf{\Sigma}_{W}\nabla g\). The
between-cluster case follows identically from the \(n_{j}\)-weighted
cluster-level estimating equations, whose scores are the
individual-level scores aggregated within clusters. \(\blacksquare\)

\textbf{Software and reproducibility.} The framework is implemented in
the developing R package \texttt{mlmediate}. The arXiv ancillary files
contain the ECLS-K analysis script, derived bootstrap summaries, and the
base-R figure-generation script. Package and additional development
materials are maintained through the causalfragility-lab GitHub
organization (\url{https://github.com/causalfragility-lab}). The
ECLS-K:2011 public-use data are distributed by the National Center for
Education Statistics and are not redistributed with the code.

\clearpage
\section{References}\label{references}

Angrist, J. D., \& Pischke, J. S. (2009). \emph{Mostly harmless
econometrics: An empiricist's companion}. Princeton University Press.

Baron, R. M., \& Kenny, D. A. (1986). The moderator--mediator variable
distinction in social psychological research. \emph{Journal of
Personality and Social Psychology, 51}(6), 1173--1182.

Bauer, D. J., Preacher, K. J., \& Gil, K. M. (2006). Conceptualizing and
testing random indirect effects and moderated mediation in multilevel
models. \emph{Psychological Methods, 11}(2), 142--163.

Degtiar, I., \& Rose, S. (2023). A review of generalizability and
transportability. \emph{Annual Review of Statistics and Its Application,
10}, 501--524.

Frank, K. A. (2000). Impact of a confounding variable on the inference
of a regression coefficient. \emph{Sociological Methods \& Research,
29}(2), 147--194.

Imai, K., Keele, L., \& Tingley, D. (2010). A general approach to causal
mediation analysis. \emph{Psychological Methods, 15}(4), 309--334.

Kahan, B. C., Li, F., Copas, A. J., \& Harhay, M. O. (2023). Estimands
in cluster-randomized trials: Choosing analyses that answer the right
question. \emph{International Journal of Epidemiology, 52}(1), 107--118.

Kenny, D. A., Korchmaros, J. D., \& Bolger, N. (2003). Lower level
mediation in multilevel models. \emph{Psychological Methods, 8}(2),
115--128.

Krull, J. L., \& MacKinnon, D. P. (2001). Multilevel modeling of
individual and group level mediated effects. \emph{Multivariate
Behavioral Research, 36}(2), 249--277.

Lüdtke, O., Marsh, H. W., Robitzsch, A., \& Trautwein, U. (2011). A $2 \times 2$
taxonomy of multilevel latent contextual models: Accuracy--bias
trade-offs in full and partial error correction models.
\emph{Psychological Methods, 16}(4), 444--467.

Lüdtke, O., Marsh, H. W., Robitzsch, A., Trautwein, U., Asparouhov, T.,
\& Muthén, B. (2008). The multilevel latent covariate model: A new, more
reliable approach to group-level effects in contextual studies.
\emph{Psychological Methods, 13}(3), 203--229.

MacKinnon, D. P. (2008). \emph{Introduction to statistical mediation
analysis}. Lawrence Erlbaum Associates.

Mundlak, Y. (1978). On the pooling of time series and cross section
data. \emph{Econometrica, 46}(1), 69--85.

National Center for Education Statistics. (2019). \emph{ECLS-K:2011
public-use kindergarten--fifth grade data file and electronic codebook}
(NCES 2019-050) [Data set]. U.S. Department of Education, Institute of
Education Sciences.

Nguyen, T. Q., Schmid, I., \& Stuart, E. A. (2021). Clarifying causal
mediation analysis for the applied researcher: Defining effects based on
what we want to learn. \emph{Psychological Methods, 26}(2), 255--271.

Preacher, K. J., Zhang, Z., \& Zyphur, M. J. (2011). Alternative methods
for assessing mediation in multilevel data: The advantages of multilevel
SEM. \emph{Structural Equation Modeling, 18}(2), 161--182.

Preacher, K. J., Zyphur, M. J., \& Zhang, Z. (2010). A general
multilevel SEM framework for assessing multilevel mediation.
\emph{Psychological Methods, 15}(3), 209--233.

Pustejovsky, J. E., \& Tipton, E. (2018). Small-sample methods for
cluster-robust variance estimation and hypothesis testing in fixed
effects models. \emph{Journal of Business \& Economic Statistics,
36}(4), 672--683.

Raudenbush, S. W., \& Bryk, A. S. (2002). \emph{Hierarchical linear
models: Applications and data analysis methods} (2nd ed.). Sage.

Robinson, W. S. (1950). Ecological correlations and the behavior of
individuals. \emph{American Sociological Review, 15}(3), 351--357.

Seaman, S. R., Pavlou, M., \& Copas, A. J. (2014). Methods for
observed-cluster inference when cluster size is informative: A review
and clarifications. \emph{Biometrics, 70}(2), 449--456.

Sobel, M. E. (1982). Asymptotic confidence intervals for indirect
effects in structural equation models. \emph{Sociological Methodology,
13}, 290--312.

Tourangeau, K., Nord, C., Lê, T., Wallner-Allen, K., Vaden-Kiernan, N.,
Blaker, L., \& Najarian, M. (2019). \emph{Early Childhood Longitudinal
Study, Kindergarten Class of 2010--11 (ECLS-K:2011): User's manual for
the ECLS-K:2011 kindergarten--fifth grade data file and electronic
codebook, public version} (NCES 2019-051). U.S. Department of Education,
National Center for Education Statistics.

VanderWeele, T. J. (2015). \emph{Explanation in causal inference:
Methods for mediation and interaction}. Oxford University Press.

VanderWeele, T. J., Hong, G., Jones, S. M., \& Brown, J. L. (2013).
Mediation and spillover effects in group-randomized trials: A case study
of the 4Rs educational intervention. \emph{Journal of the American
Statistical Association, 108}(502), 469--482.

Vansteelandt, S., \& Daniel, R. M. (2017). Interventional effects for
mediation analysis with multiple mediators. \emph{Epidemiology, 28}(2),
258--265.

Zhang, Z., Zyphur, M. J., \& Preacher, K. J. (2009). Testing multilevel
mediation using hierarchical linear models: Problems and solutions.
\emph{Organizational Research Methods, 12}(4), 695--719.

\end{document}